\documentclass[twocolumn]{aastex631}
\newcommand{\unit}[1]{\ \mathrm{#1}}

\newcommand{\di}[3][]{\frac{d^{#1} #2}{{d #3}^{#1}}}
\newcommand{\pa}[3][]{\frac{\partial^{#1} #2}{{\partial #3}^{#1}}}
\newcommand{\Msol}{M_{\odot}}
\newcommand{\ord}[1]{\times 10^{#1}}
\newcommand{\vectorb}[1]{\mbox{\boldmath $#1$}}
\newcommand{\average}[1]{\ensuremath{\langle#1\rangle} } 

\newcommand{\enc}[1]{\left(#1\right)}
\newcommand{\encl}[1]{\left[#1\right]}
\definecolor{forest-green}{rgb}{0.13,0.55,0.13}

\accepted{March 20, 2024}

\submitjournal{ApJ}

\shorttitle{MHD simulation in the inner Galaxy with thermal processes}
\shortauthors{Kakiuchi et al.}

\usepackage{amsmath,amssymb}
\usepackage{tabularx}

\begin{document}

\title{MHD Simulation in Galactic Center Region with Radiative Cooling and Heating}

\correspondingauthor{Kensuke Kakiuchi}
\email{kakiuchi@g.ecc.u-tokyo.ac.jp, kakiuchi.mw@gmail.com}
\author[0009-0006-8097-8966]{Kensuke Kakiuchi}
\affiliation{School of Arts \& Sciences, The University of Tokyo, 3-8-1, Komaba, Meguro, Tokyo 153-8902, Japan}
\affil{Dept. of Physics, Nagoya University, Nagoya, Aichi 464-8692, Japan}

\author{Takeru. K. Suzuki}
\affiliation{School of Arts \& Sciences, The University of Tokyo, 3-8-1, Komaba, Meguro, Tokyo 153-8902, Japan}

\author{Shu-ichiro Inutsuka}
\affiliation{Dept. of Physics, Nagoya University, Nagoya, Aichi 464-8692, Japan}

\author{Tsuyoshi Inoue}
\affiliation{Department of Physics, Konan University, Okamoto 8-9-1, Higashinada-ku, Kobe, Hyogo 658-8501, Japan}

\author{Jiro Shimoda}
\affiliation{Institute for Cosmic Ray Research, The University of Tokyo, 5-1-5, Kashiwanoha, Kashiwa, Chiba 277-8582, Japan}




\begin{abstract}

We investigate the role of magnetic field on the gas dynamics in {a g}alactic bulge region by three dimensional simulations with radiative cooling and heating. While high-temperature corona with $T>10^6\unit{K}$ is formed in the halo regions, the temperature near the {mid-plane} is $\lesssim 10^4\unit{K}$ following the thermal equilibrium curve determined by the radiative cooling and heating. Although the thermal energy of the interstellar gas is lost by radiative cooling, the saturation level of the magnetic field strength does not significantly depend on the radiative cooling and heating. The magnetic field strength is amplified to $10\unit{\mu G}$ on average, and reaches several hundred ${\rm \mu G}$ locally. We find the formation of magnetically dominated regions at mid-latitudes in the case with the radiative cooling and heating, which is not seen in the case without radiative effect.
The vertical thickness of the mid-latitude regions is $50-150 \unit{pc}$ at the radial location of $0.4-0.8 \unit{kpc}$ from the {g}alactic center, which is comparable to the observed vertical distribution of neutral atomic gas.
When we take the average of different components of energy density integrated over the {g}alactic bulge region, the magnetic energy is comparable to the thermal energy.
We conclude that the magnetic field plays a substantial role in controlling the dynamical and thermal properties of the {g}alactic bulge region.

\end{abstract}




\section{Introduction} \label{sec:intro}

The {neutral atomic hydrogen (${\rm H}_{\rm I}$)} gas in the Galactic bulge (GB hereafter) region ($R <1.0$ kpc) has a thickness of over 70 pc in the vertical direction \citep{Sofue22c,Sofue22d}.
On the other hand, if we assume the {force} balance, the vertical scale height can be estimated to be $h =c_s/\Omega \approx 20$ pc\footnote{{In a strict sense $\sqrt{\pa{g_z}{z}}$ should be used for $\Omega$ where the $g_z$ is vertical component of gravity (see Section \ref{sec:thickness-beta}).}}
, where $c_s \approx 10$ km s$^{-1}$ is the sound velocity in the gas with temperature $\approx 10^4$ K and $g_z=-\pa{\Phi}{z}$, is the observed rotational frequency \citep[e.g.,][]{Sofue13a}.
$c_s$ could be smaller if the temperature of ${\rm H}_{\rm I}$ gas is lower, which gives further smaller $h$.
This comparison indicates that the the vertical distribution of the ${\rm H}_{\rm I}$ gas cannot be supported only by the gas pressure, suggesting the presence of additional mechanisms required to maintain the observed vertical thickness.

The magnetic field is one of such plausible mechanisms; the vertically extended gas can be explained by the contribution from magnetic pressure in combination with gas pressure.
In addition, 
there have been a variety of complex structures observed that are expected to reflect magnetic-field configurations, such as molecular loops\footnote{The non-axisymmetric potential of the stellar bar is also discussed as an alternative mechanism to explain the loop-like velocity pattern \citep{Liszt06a,Liszt08a,Sormani19c}.}
\citep{Fukui06a,Fujishita2009a,Torii10a,Torii10b}, spiral-shaped molecular gas \citep[the Double Helix Nebula;][]{Morris2006a,Enokiya2014a} and non-thermal {filamentary structures} \citep{Yusef-Zadeh84a,Morris1985a,Anantharamaiah1991a,Lang1999b,LaRosa2004a,Pare19a,Heywood22a,Yusef-Zadeh2022b}.
{These observations illustrate that the magnetic field plays a significant role in GB region.}
{Similar magnetic properties are obtained in other spiral galaxies \citep{Golla1994a,Sofue1994a,Konishi2022a}.}

The magnetic field strength {in the GB region of the Milky way} has been estimated by multiple independent observational techniques.
For instance, 
{if the shape of the Radio Arc{, the most extended filamentary structures in the GB region,} is maintained by magnetic fields in surrounding turbulent medium, the field strength of $B=0.1-1 \unit{mG}$ is required} \citep{Morris96a}.
\citet{Pillai15a} observed the polarization of the dark nebula G0.253+0.016 (known as `Brick') and estimated the magnetic field strength to be a few mG using the Davis-Chandrasekhar-Fermi (DCF) method \citep{Davis1951a,Chandrasekhar53a}.
An upper limit of the nondetected $\gamma$-rays by EGRET gives a lower limit on the average magnetic field strength $\ge 0.05$ mG within 400 pc of the Galactic center region \citep{Crocker10a}; the magnetic field needs to be sufficiently strong in order that cosmic-ray electrons effectively lose their energy via Synchrotron radiation to keep the inverse Compton radiation less than the detection limit.
On the basis of these {estimates}, the energy density of the magnetic field equals to or even exceeds the thermal energy density of the gas in the GB region. 
{Such strong magnetic fields are also reported in the central region of other spiral galaxies \citep[e.g.,][]{Beck2005a,Tabatabaei2018a}.}
Thus, it is essential to take into account magnetohydrodynamical (MHD) effects in order to understand the properties of the gas in the GB region {of spiral galaxies.}

It is also inferred that magnetic fields may affect star formation activity.
While there {is molecular cloud complex} called the Central Molecular Zone (CMZ) in the Milky way, the star formation rate, $\sim0.02–0.1 \unit{M_\odot\ yr}^{-1}$ \citep[e.g.][]{Yusef-Zadeh2009a, Immer2012a}, {integrated over the GB region} is much smaller than that estimated by extrapolation from the {G}alactic disk \citep{Longmore2013a, Kruijssen2014a}. 
{The reduced star formation may be a consequence of {strong} magnetic activity \citep{Morris1993a,Chuss2003,Henshaw2022a}}

Numerical investigations on the effect of magnetic fields in the dynamics of the GB region have also been carried out.
For example, global MHD simulations by \citet{Machida09a} demonstrated that gas is lifted up from the {mid-plane} by buoyantly rasing magnetic loops, which is a possible mechanism for the observed molecular loops \citep{Fukui06a}.
\citet{Suzuki15a} reported with a global MHD simulation that the observed parallelogram on the {position-velocity} diagram can be reproduced by radial flows excited by a non-smooth radial profile of rotational velocity.
{In the same simulation, ``sliding slopes'' associated with magnetic loops are also ubiquitously found in a time-dependent manner, which drive high-speed downflows in excess of 100 km s$^{-1}$ \citep{Kakiuchi18a}.}
These numerical simulations treat only high-temperature gas with $T\sim10^{6}$ K to avoid {numerical difficulty in the description of low-temperature fine-scale clouds.}
However, it is essentially important to handle lower-temperature gas with radiative cooling in order to model the realistic interstellar condition.
In this paper, we investigate the magnetic activity in the {GB region} of {spiral galaxies} in detail, taking into account the thermal evolution including 
{radiative cooling and heating in} the numerical simulation code of \citet{Suzuki15a}
\footnote{{Recently,} {\citet{Moon23a} run MHD simulations with stellar activity and radiative cooling, focusing on formation of nuclear rings by magnetised inflows.}}
.

\section{Numerical Method} \label{sec:method}
We perform three dimensional (3D) global MHD simulations to study the time evolution of the gas disk in the GB region
with an initially weak vertical magnetic field. 
Our simulations are performed in a cylindrical coordinate system ($R$, $\phi$, $z$).
{
In this work, we adopt an axisymmetric gravitational potential.
We will present an explanation of the numerical method in the following order: basic equations (Section \ref{sec:basic_equations}), the gravitational potential (Section \ref{sec:potential}), radiative cooling and heating (Section \ref{sec:cooling_and_heating}), and numerical setup including initial and boundary conditions (Section \ref{sec:setup}).
}

\subsection{Basic equations}\label{sec:basic_equations}

We solve the time evolution of the gas with the following ideal MHD equations 
{with the second-order Godunov CMoC-CT method \citep{Evans-Hawley1988a,Clarke1996a,Sano1999,Suzuki14a,Suzuki15a}}
including {radiative cooling and heating}:

{
\footnotesize
\begin{gather}
    \frac { \partial \rho } { \partial t } + \nabla \cdot ( \rho \boldsymbol { v } ) = 0,
    \\
	\rho \frac { \partial \boldsymbol { v } } { \partial t } = - \rho ( \boldsymbol { v } \cdot \nabla ) \boldsymbol { v } - \nabla \left( P_{\rm g} + \frac { B ^ { 2 } } { 8 \pi } \right) + \left( \frac { \boldsymbol { B } } { 4 \pi } \cdot \nabla \right) \boldsymbol { B } - \rho \nabla \Phi,
	\label{eqn_momentum}
    \\
	\frac { \partial \vectorb { B } } { \partial t } = \nabla \times ( \boldsymbol { v } \times \boldsymbol { B } ),\label{eqn_induction}
	\\
	\frac { \partial } { \partial t } \left[ \rho \left( e + \frac { v ^ { 2 } } { 2 } + \Phi \right) + \frac { B ^ { 2 } } { 8 \pi } \right] \nonumber
	\\+ \nabla \cdot \left[ ( \rho \boldsymbol { v } ) \left( e + \frac { P_{\rm g} } { \rho } + \frac { v ^ { 2 } } { 2 } + \Phi \right) + \frac { c } { 4 \pi } ( \boldsymbol { E } \times \boldsymbol { B } ) \right] = -{\enc{\frac{\rho}{\mu \mathrm{m_H}}}}\mathcal { L }.\label{eqn_energy}
\end{gather}
}
%
We assume an equation of state for ideal gas,
\begin{equation}
    P_{\rm g}=(\gamma-1)\rho e,
    \label{eqn_state}    
\end{equation}
where $\gamma = 5/3$ is the ratio of specific heats. 
The thermal energy per mass $e$ is expressed as
$e = \frac{\textrm{k}_\textrm{b}T}{(\gamma-1)\mu \textrm{m}_\textrm{H}}$
{where $T$ is temperature,} $\textrm{k}_\textrm{b}$ is the Bolzman constant, $\textrm{m}_\textrm{H}$ is the unified atomic mass unit.
\begin{figure}
    \centering
    \plotone{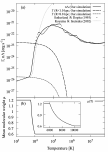}
    \caption{(a) The {cooling and heating rates against} temperature.
    The solid black line denotes the cooling {rate for} $n=1.0$ cm$^{-3}$ and the dash black lines denote the heating function with $G(R)=10^3\textrm{G}_0$ ({in the GB region $R<1.0$ kpc) and} $\textrm{G}_0$ ({at the {solar} neighborhood}) 
    in our simulation. {The black dots plot raw data by \citet{Sutherland93a} and the gray dotted line plots the fitting function by \citet{Koyama02a}.}
    (b) The mean molecular weight adopted in our simulation.
    }
    \label{fig:ini_T2mu}
\end{figure}
We adopt the following approximated formula for the mean molecular weight{, $\mu$,} that depends on temperature, 
\begin{align}
    \mu(T) = 
    \begin{cases}
        {1.2\ (T<6000 \unit{K})}\\
        {\frac{0.342}{1-{4289.4}{T}^{-1}}\ (6000 \unit{K} \leq T < 10000 \unit{K}) }\\
        {0.6\ (T\geq 10000 \unit{K})}
    \end{cases}
    ,
\end{align}
in order to take into account the effect of the ionization, which is presented in Figure \ref{fig:ini_T2mu}(b).
$\Phi$ {in equation (\ref{eqn_momentum})} is the gravitational potential, which will be described in Section \ref{sec:potential}.
{{We do not consider the self-gravity of the gas.}}
$\mathcal{ L }$ {in equation (\ref{eqn_energy})} represents the net cooling rate [erg s$^{-1}$] as a function of number density $n=\rho/ \mu m_H$ and temperature $T$:
\begin{align}
	\mathcal{ L }=n\Lambda(T) - \Gamma(T),
	\label{eqn:cooling_and_heating}
\end{align}
where {$\Lambda$ [erg s$^{-1}$ cm$^{3}$] is cooling efficiency and $\Gamma$ [erg s$^{-1}$]} is heating rate.
Figure \ref{fig:ini_T2mu}(a) shows the cooling and heating function{s} for temperature, which will be described in Section \ref{sec:cooling_and_heating}.

\subsection{The gravitational potential}\label{sec:potential}
We perform our MHD simulations under time-independent and axisymmetric Galactic gravitational potential. This potential consists of four components of gravitational sources, the super-massive black hole (SMBH) at the Galactic center, the stellar bulge, the stellar disk, and the dark matter halo; 
\begin{align}
    \Phi=\Phi_\mathrm{BH}+\Phi_\mathrm{*,bulge}+\Phi_\mathrm{*,disk} +\Phi_\mathrm{halo}. \label{eqn_gravity_all}
\end{align}
We assume that the SMBH as a point mass of $4.4\ord{6}\Msol$\citep{Genzel10a}.
The gravitational potential of the stellar bulge ($i=1$) and disk ($i=2$) is adopted from \citet{Miyamoto75a};
\begin{align}
	\Phi_{*}=\sum_{i=1}^{2}{\frac{\mathbf{-}\mathrm{G}M_i}{\sqrt{R^2+(a_i+\sqrt{z^2+b_i^2})^2}}},
	\label{eqn_gravity}
\end{align}
where G is the gravitational constant.
The fit parameters are set to be $M_1=2.05\ord{10}\Msol$, $M_2=2.57\ord{11}\Msol$, $a_1=0.0$, $a_2=7.258$, $b_1=0.495$ and $b_2=0.52$, respectively.

For the dark matter halo, we adopt the NFW model \citep{Navarro96a,Sofue13a}:
\begin{align}
    \Phi_\mathrm{halo}=-4\pi \mathrm{G} \rho_\mathrm{h,0} r_\mathrm{h}^3
    \ \frac{1+r/r_\mathrm{h}}{r},
\end{align}
where $r = \sqrt{R^2+z^2}$ is the spherical radius.
We assume $r_\mathrm{h}=10.7$ kpc and $\rho_\mathrm{h,0}=1.82\ord{-2} \Msol \unit{pc}^{-3}$.

\subsection{Cooling and heating} \label{sec:cooling_and_heating}

We adopt the radiative cooling of the solar-metallicity gas for simplicity, {whereas} the metallicity gradually increases toward the center to give a few times greater value in the {GB} region than in the solar neighborhood \citep{Shields94a,Maeda02a}.
We cover the temperature range of $10^3 \unit{K}<T<7\ord{7}\unit{K}$.
For the non/weakly ionized gas with $T<10^4\unit{K}$, we adopt the cooling function $\Lambda_{\rm l}$ for the interstellar medium by \citet{Koyama02a}, which is shown by the gray dashed line in Figure \ref{fig:ini_T2mu}(a).
For the (almost) fully ionized gas with $T>10^4\ \unit{K}$,
we prescribe the cooling function $\Lambda_{\rm h}$ for optically thin plasma by \citet{Sutherland93a}.
A fitting formula for $\Lambda_{\rm h}$ is presented in 
Appendix \ref{sec:fitting_lamda}. 
We smoothly connect $\Lambda_{\rm l}$ and $\Lambda_{\rm h}$ via
\begin{align}
    \Lambda&=0.5\encl{\Lambda_{\rm l}(1-f)+\Lambda_{\rm h}(1+f)},
\end{align}
with
\begin{align}
    f&=\tanh\enc{{\frac{\log_{10}T-\log_{10}T_\textrm{b}}{\log_{10}(\Delta T_\textrm{b})}}},
\end{align}
where $T_\textrm{b}=10^4$ K, $\log_{10}(\Delta T_\textrm{b})=0.1$. 
The first term of equation (\ref{eqn:cooling_and_heating}),
$n \Lambda$, for $n=1$ cm$^{-3}$ is shown as the black solid line in Figure \ref{fig:ini_T2mu}.

We do not solve low-temperature ($T<10^3 \unit{K}$) clouds, because our simulations focus on the global properties of the GB region. However, these should be taken into account in more elaborated simulations to study star formation activity because they contain large mass in a small volume. Their scale height is typically $\lesssim$ several pc, which is comparable to the minimum resolution, a few pc, in the current simulation setup; higher-resolution is required to capture such dense cool clouds.
We discuss this topic in Section \ref{sec:discuss_CNM}.

The heating rate from the interstellar radiation field (ISRF) depends on the surrounding stellar environment, such as the OB association.
Therefore, the radiation heating rate is relatively high near the Galactic center where the stellar density is higher than {in} the solar neighborhood \citep{Wolfire03a}.
The ISRF first heats up dust grains. The gas is eventually heated through collisions with the dust grains. We here assume that the dust and gas components are thermally well coupled via sufficiently frequent collisions.
The temperature is set by the balance between the heating due to the absorption of photons from the ISRF and the cooling due to thermal radiation from the dust particles.

\begin{figure}
    \centering
    \plotone{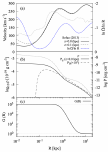}
    \caption{Initial distribution for Galactic distance $R$.
    (a) Velocity and strength of differential rotation.
    (b) Density, gas pressure and magnetic pressure. (c) $\mathcal{G}(R)$.}
    \label{fig:initial_condition}
\end{figure}

The ISRF intensity is given as a local profile, $\mathcal{G}(R)$, relative to 
the standard value, {$\mathcal{G}_0$}, calibrated at the solar neighborhood \citep{Draine78a}.
The ISRF is empirically estimated to be 1000 $\mathcal{G}_0$ \citep[e.g.,][]{Clark13a}; this value reproduces the observed thermal properties of 'Brick'. \citet{Sormani18a} introduced a functional form for $\mathcal{G}(R)$ to mimic the empirically derived profile (Figure \ref{fig:initial_condition}(c)), which we adopt in our simulations. 
The heating rate $\Gamma(R)$ is treated in our simulations as proportional to $\mathcal{G}(R)$.


The heating rate due to cosmic-rays is scaled to the ionization rate by cosmic-rays $\xi_\mathrm{CR}$.
In the solar neighborhood, $\xi_\mathrm{CR}=3.0\ord{-17}$ s$^{-1}$.
As with the ISRF, the ionization rate near the Galactic center is higher than one at the solar neighborhood because of the increase in the flux of cosmic-rays. The ionization rate is estimated to be $\xi_\mathrm{CR}\sim 2.0\ord{-14}$ s$^{-1}$
{in the inner} several 100 parsec \citep{Oka19a}, higher than in the solar neighborhood by {a} factors of {$\sim 10^3$}.
Then, we assume that the ionization rate, and accordingly the heating rate by cosmic-rays, is also simply proportional to $\mathcal{G}(R)$ as in the ISRF.

We take account of the heating from ISRF and cosmic-rays, and model the function of heating rate in our simulations {as follows}:
\begin{align}
    \Gamma(R,T) = \Gamma_0\mathcal{G}(R)\exp\enc{-\frac{T}{T_{\rm cut}}}, \label{eqn_ht-Gamma}
\end{align}
where $\Gamma_0=2.0\ord{-26} \unit{erg\ s}^{-1}$ \citep{Koyama02a} is the heating rate at the solar neighborhood.
The heating from the ISRF is ineffective above a certain temperature 
{because the intensity of the UV radiation is mainly determined by massive stars.
We set $T_{\rm cut}=5.0\ord{4}$ K, referring to the effective temperature of O-type stars.
In reality, the cosmic-ray heating is still active in the higher-temperature regime. However, we ignore the heating above $T_{\rm cut}$ for simplicity, because the bulk of the heating is dominated by the UV radiation from massive stars.}
Heating functions, $\Gamma$, for $\mathcal{G}(R)=10^3\mathcal{G}_0$ in $R < 1$ kpc (long dashed line) and $\mathcal{G}(R) = \mathcal{G}_0$ at the solar neighborhood (short dashed line) are compared in Figure \ref{fig:ini_T2mu}.
%

\begin{figure}
    \centering
    \plotone{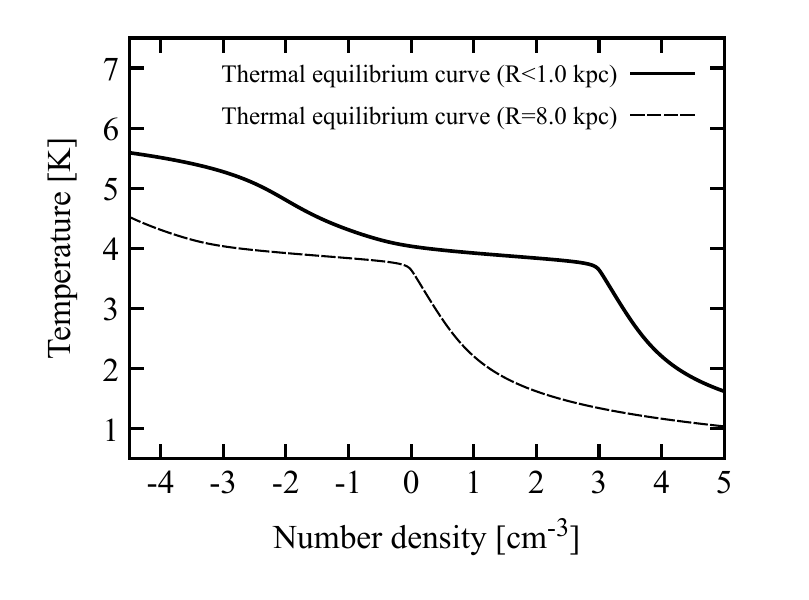}
    \caption{The thermal equilibrium curves in the $n-T$ diagram when $\mathcal{G}(R)=10^3 \mathcal{G}_0$ ($R<1.0$ kpc; solid line), $\mathcal{G}_0$ ($R=8.0$ kpc; dash line).}
    \label{fig:n2T_diagram}
\end{figure}

The temperature of the gas in thermal equilibrium can be derived from $\mathcal{ L }=0$ 
{when the heating is balanced with the cooling.}
Figure \ref{fig:n2T_diagram} shows thermal equilibrium curves in an $n-T$ diagram in $R < 1.0$ kpc (solid) and at {the solar neighborhood} (dashed).

\subsection{Numerical Setup} \label{sec:setup}

The initial gas distribution satisfies {the equilibrium configuration; the force balance among the gravity, the pressure gradient force and the centrifugal force are fulfilled}:
\begin{align}
    -\frac{1}{\rho}\pa{P_{\rm g}}{R}+\frac{v_\phi^2}{R}+g_R=0, \label{eqn_forcebalnce_R}\\
    -\frac{1}{\rho}\pa{P_{\rm g}}{z}+g_z=0 \label{eqn_forcebalnce_z},
\end{align}
where $g_R=-\pa{\Phi}{R}$ and $g_z=-\pa{\Phi}{z}$.
{We present a specific method to determine the initial conditions in appendix \ref{sec:Appendix_B}.}

Figure \ref{fig:initial_condition} (a) and (b) show the radial distributions of the initial rotational speed, density, and pressure.
We set up a weak vertical magnetic field initially
with the following radial dependence:
\begin{align}
	B_z= 0.7 \enc{\frac{R}{1\ \unit{kpc}}}^{-1}\exp{\enc{-\frac{0.1}{R-R_b}} } \unit{\mu G}, 
    \label{eqn_initial_magnetic_field}	
\end{align}
where $R_b=0.08$ kpc and set to $B_z=0$ in $R < R_b$.

\begin{table}[t]
    \centering
    \begin{tabularx}{0.82\linewidth}{c|c|c}
        \hline
        $R$ (kpc) & $\phi$ &$z$ (kpc)\\
        \hline
        $0.01<R<50$ & $-\pi<\phi<\pi$ & $-0.5 < z <0.5$ \\
        \hline
    \end{tabularx}
    \caption{Simulation domain}
    \label{table:mesh}
\end{table}

\begin{table*}[t]
\centering
\begin{tabularx}{0.6\linewidth}{c||c|c|c|c|c}
\hline
Case & $N_R$ & $N_\phi$ &$N_z$ & Radiative cooling and heating &
Simulation time 
\\
\hline
I& 384 & 256 & 320 & Off & 250 Myr \\
II& 384 & 256 & 320 & Fiducial & 204 Myr \\
I-l & 192 & 128 & 160 & Off & 250 Myr \\
II-l & 192 & 128 & 160 & Fiducial & 225 Myr \\
III-l & 192 & 128 & 160 & Weak & 250 Myr \\
IV-l & 192 & 128 & 160 & Strong & 250 Myr \\
\hline
\end{tabularx}
\caption{{Summary of simulation cases.}}
\label{table:case}
\end{table*}

The simulation domain covers the full $2\pi$ in azimuth. The radial and vertical domain sizes are $0.01 \unit{kpc} < R < 50 \unit{kpc}$ and $-0.5 \unit{kpc} < z < 0.5 \unit{kpc}$, {respectively} (Table \ref{table:mesh}). {The outflow condition is employed for the $R$ and $z$ boundaries} \citep[e.g.,][]{Suzuki14a}.
At the inner {$R$ boundary}, $B_z=0$ is set for the vertical magnetic field.


Table \ref{table:case} presents four cases of numerical simulations conducted for the current paper. Case II is the fiducial case. In Case I the radiative cooling and heating are turned off but the numerical resolution and other setup are the same as those in Case II. By comparison between Cases I and II, we investigate the role of the radiative effects.
We reduce (enhance) the net cooling in Cases III (IV) -l by changing the radiative heating rate. In Cases III-l and IV-l lower numerical resolution is employed to save computational time. 
{We also perform half-resolution simulations with the same setup as in Cases I and II, which are labeled as Cases I-l and II-l, in order to directly compare to Cases III-l and IV-l.}

The sizes of radial and vertical cells, $\Delta R$ and $\Delta z$, are enlarged in proportion to $R$ and $z$, respectively. In Cases I and II, the minimum cell size is $\Delta R \approx \Delta z \approx 1$ pc.  We run the simulations after the magnetic field strength is sufficiently saturated in a quasi-steady manner. The simulations times are summarized in Table \ref{table:case}.

\section{Results}


\subsection{Overview}\label{sec:overview}

\begin{figure*}[ht!]
    \centering
    \plotone{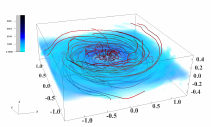}
    \caption{
    3D overview inner the GB region at $t=$200.1 Myr {in {C}ase II.}
   Colors represent number density. 
   Red lines represent magnetic field lines. 
   }
    \label{fig:overview}
\end{figure*}

Figure \ref{fig:overview} {shows a 3D snapshot} in $R<1.5$ kpc at $t=$200.1 Myr in Case II.
One can see that the overall magnetic structure is mostly dominated by the toroidal component as a result of the winding due to differential rotation.
On the other hand, in the denser central region of $R < 300$ pc, one can see vertically extended poloidal magnetic fields.
{We examine in detail the numerical results of the case with the radiative cooling and heating in comparison to those with {adiabatic} setting below.}

\subsubsection{Density distribution}\label{sec:density}
\begin{figure*}
    \centering
    \plottwo{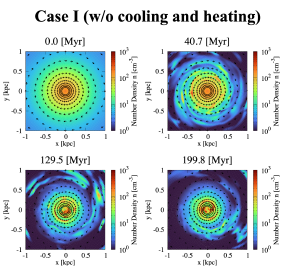}{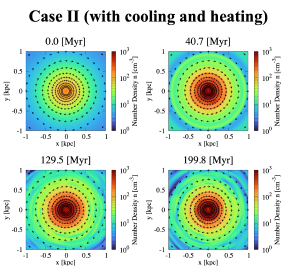}
    \caption{
    Time evolution of the number density profile on the {mid-plane} ($z = 0.0$ kpc). The left {four} panels show the results for Case I (without radiative cooling and heating) and the right {four} panels show the results for Case II (with radiative cooling and heating). The snapshots are taken at $t =$ 0.0, 40, 100, and 200 Myr. The arrows represent velocity vectors in the {mid-plane}.
    }
    \label{fig:faceon}
\end{figure*}

\begin{figure}
    \centering
    \plotone{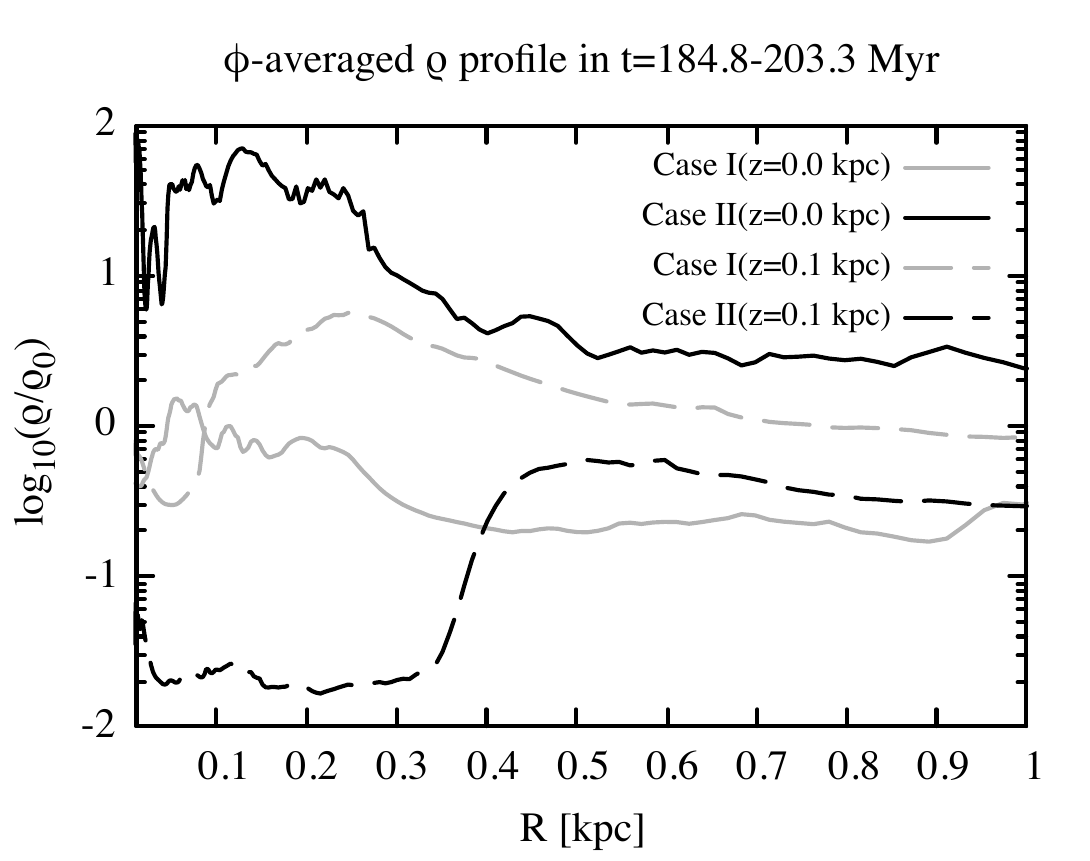}
    \caption{
    Comparison of radial distributions of the mass density 
    averaged over time, $t=184.8 - 203.3 \unit{Myr}$ and $\phi$ direction, normalized by the initial mass density. The black and gray lines indicate the results with cooling and heating (Case II) and without cooling and heating (Case I), respectively. The solid and dotted lines are 
    at $z = 0.0 \unit{kpc}$ and $0.1 \unit{kpc}$, respectively.
    }
    \label{fig:density_profile_in_radial_profile}
\end{figure}

We show the time evolution of the number density profiles on the {mid-plane} ($z = 0.0$ kpc) in Figure \ref{fig:faceon} at $t =$ 0.0 (top left), 40 (top right), 130 (bottom left), and 200 (bottom right) Myr {of the two cases}.
We can see that {the radiative cooling and heating substantially affect the density distribution on the {{mid-plane}}.}
In Case I {without radiative effects (left of Figure \ref{fig:faceon})}, the gas density is considerably decreasing from the initial distribution over time.
We can also confirm the formation of clear non-axisymmetric structure.
By contrast, the density increases rapidly at an early stage ($t=40$ Myr) in Case II {with radiative effects (right of Figure \ref{fig:faceon})}. 
In particular, it {increases} more dramatically in the inner region.

Figure \ref{fig:density_profile_in_radial_profile} shows the radial distribution of mass density, $\rho$, normalized by the initial {value}, $\rho_0$, within $R<1$ kpc, where $\rho$ is spatially averaged over full azimuthal angle and over time $t = 184.4 - 203.3 \unit{Myr}$. 
The gray and black lines show the results of Cases I and II, respectively.
The solid (dotted) line depicts the distribution at $z=0.0\ (0.1)$ kpc.
Comparison between Cases I and II clearly shows the importance of the radiative cooling and heating. In Case I without radiative effects, the density on the {mid-plane} is decreasing ($\rho/\rho_0<1.0$) but the density above the {mid-plane} is increasing ($\rho/\rho_0>1.0$).
This is because the gas expands vertically {as a consequence of magnetic processes (Section \ref{sec:magnetic_heating})}
and mass is supplied from the {mid-plane} to the upper regions.
On the other hand, {in Case II with radiative cooling and heating,} the density on the {mid-plane} is increasing, while the density 
at $z=0.1$ kpc is decreasing. This is a consequence of the downflows from upper regions to the {mid-plane}. 
The temperature near the {mid-plane} drops rapidly by efficient radiative cooling after the simulation starts because of the high density, which causes the decrease in the gas pressure. As a result, {the dynamical equilibrium in the vertical direction} is no longer satisfied, which induces the downflows. We explain the details in section \ref{sec:poloidal_structure}.

\subsubsection{Amplification of the magnetic field}\label{sec:tevol_B}
\begin{figure}
    \centering
    \plotone{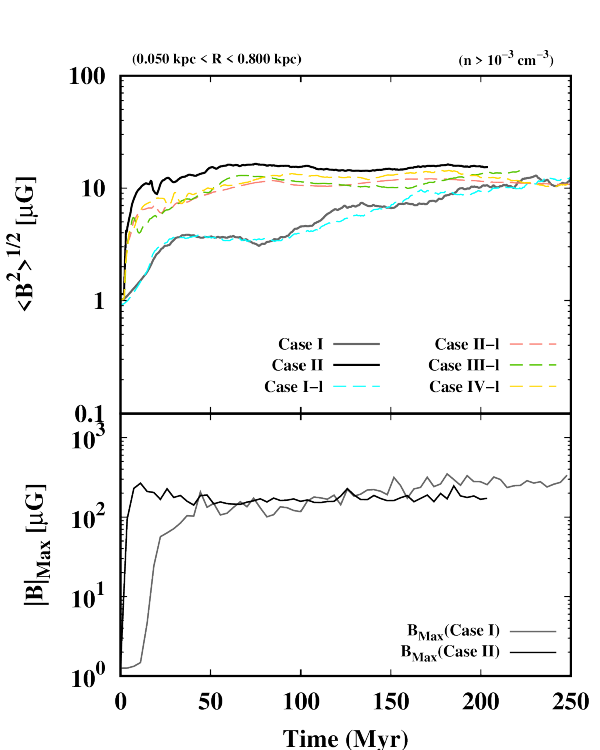}
    \caption{{Time evolution of the volume-averaged magnetic field strengths of various cases (top) and the maximum field strengths in the same region for two cases (bottom).}}
    \label{fig:tevol_B}
\end{figure}

In these simulations, the {initially} weak vertical magnetic field {is} amplified by various mechanisms; one of the reliable processes is
magnetorotational instability 
\citep[hereafter MRI;][]{Velikhov1959,Chandrasekhar61a,Balbus91a}.
MRI occurs in inner-fast differentially rotating systems, and {small radial fluctuations of} magnetic field lines {are} amplified through the transport of the angular momentum between inner and outer rings of a disk.
The differential rotation also causes winding of poloidal magnetic field lines, resulting in the amplification of $B_{\phi}$.

In Figure \ref{fig:tevol_B}, we show the time evolution of the 
{spatially averaged root mean squared (r.m.s.)} magnetic field strength (solid lines),
 \begin{align}
 \average{B^2}^{1/2}_{{R,\phi,z}}=
 \enc{\frac{\iiint{B^2 R\textrm{d}R \textrm{d}\phi\textrm{d}z}}
 {\iiint{R\textrm{d}R \textrm{d}\phi \textrm{d}z}}}^{1/2},    
 \end{align}
where $B^2=B_\phi^2+B_R^2+B_z^2$, and the averaged ranges are $0.05\unit{kpc}<R<0.8 \unit{kpc}$, $-\pi<\phi<\pi$ and $|z|<0.45 \unit{kpc}$.
The gray and black lines are the 
{the results without (Case I) and with (Case II) radiative cooling and heating}, respectively.
In Case II, the {average} magnetic field strength {(solid black line)} is amplified rapidly to $10\ \mu$G in a short time within $t=$10 Myr.
In Case I, the magnetic field {is amplified} relatively slowly, but after $t\gtrsim$ 200 Myr, 
{the saturated average field strength is almost comparable to that of Case II.}
The dotted lines in Figure \ref{fig:tevol_B} plot the maximum value of the magnetic field among all grid points in the same region for Case I (gray) and Case II (black), respectively.
Although the saturation times of the two cases are different, the maximum field strengths in the saturated state reach
$200-300\ \mu $G, which is similar in both cases.
{Although in the lower resolution Cases I-l and II-l the growth of the magnetic fields is slightly slower than in the high-resolution counterparts of Cases I and II, similar saturated field strengths are achieved in both low and high resolution simulations.}
\begin{figure}
    \centering
    \plotone{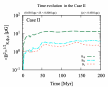}    
    \caption{Time evolution of 
    {radial (cyan dot-dashed), azimuthal ({dark-green} dashed), and vertical ({pale red} dotted) components of the magnetic field averaged over the R, $\phi$ and z directions.
    The averaged range is the same as in Figure \ref{fig:tevol_B}}.}
    \label{fig:B_compornent_t}
\end{figure}

\begin{figure}
    \centering
    \plotone{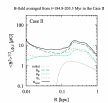}
    \caption{Radial profile of magnetic field. {The lines are the same as in Figure \ref{fig:B_compornent_t}, except for gray dotted line (initial condition).}}
    \label{fig:B_compornent_R}
\end{figure}

When the radiative cooling and heating are included (Case II), the magnetic field grows more rapidly from the beginning within the initial 10 Myr.
In order to understand this situation in more detail, the {volume averaged} total magnetic field strength $B(=\sqrt{B_z^2+B_R^2+B_{\phi}^2}$) is divided into each component, $B_z$ (red), $B_R$ (blue), and $B_{\phi}$ (green), and their time evolutions are {compared in Figure \ref{fig:B_compornent_t}.}
It is clear from this figure that the azimuthal component, $B_{\phi}$, is most strongly amplified owing to winding by the {differential} rotational from the early {time}. 
{In addition, particularly at the beginning,} the magnetic field is {also} compressed by the downflows toward the {mid-plane}, 
{which contributes to the amplification of $B_\phi$.}
The compressible amplification continues until the magnetic pressure-gradient force is balanced by the downward gravity force.

We also show {in Figure \ref{fig:B_compornent_R}} the radial distribution of the magnetic field averaged over time and the $\phi$ and $z$ directions
\begin{align}
  \average{B^2}^{1/2}_{{t,\phi,z}}=
  \enc{\frac{\iiint{B^2 \textrm{d}t \textrm{d}\phi\textrm{d}z}}
  {\iiint{\textrm{d}t \textrm{d}\phi \textrm{d}z}}}^{1/2}. 
  \end{align}
The r.m.s. $B$ (black solid line), $B_z$ (red dotted line), $B_R$ (green dash line), and $B_{\phi}$ (blue dotted dash line), are compared to the initial $B_z$ (gray dotted line). 
{The field strength, $|B|$, is amplified by more than 10 times compared to the initial vertical field strength. 
The ratios of poroidal components to toroidal component, $|B_z|/|B_\phi|$ and $|B_R|/|B_\phi|$, are 0.16 and 0.28, respectively. 
These ratios are consistent with the results obtained from previous global or semi-global simulations \citep{Brandenburg1995a,Balbus1998,Hawley2000a,Hawley2011a,Suzuki2014a,Suzuki2023a}.
}

\subsubsection{Time evolution of energy density}\label{sec:energy_density}
\begin{figure*}
    \centering
    \plottwo{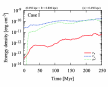}{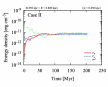}
    \caption{Time evolution of {thermal (blue dotted), magnetic (red solid),
    and turbulent (green dashed) energy densities}
    averaged over the $R$, $\phi$ and $z$ directions.}
    \label{fig:tevol_prsgas}
\end{figure*}

{
We show in Figure \ref{fig:tevol_prsgas} the time evolutions of the volume averaged different components of pressure,
\begin{align}
 \average{P}^{1/2}_{{R,\phi,z}}=
 \enc{\frac{\iiint{p\ R\textrm{d}R \textrm{d}\phi\textrm{d}z}}
 {\iiint{R\textrm{d}R \textrm{d}\phi \textrm{d}z}}}^{1/2},    
\end{align}
for Case I (left panel) and Case II (right panel). The averaged {region} is the same as in Figure \ref{fig:tevol_B}.
Each line denotes magnetic pressure, $P_B=B^2/8\pi$, (red solid), thermal pressure, $P_g=(\gamma-1)\rho e$, (dotted blue), and turbulent pressure, $\rho v^2=\rho[v_R^2+(v_{\phi}-v_{\phi,0})^2+v_z^2] $, (green dashed), respectively.
Here, the turbulent component is the kinetic energy excluding the initial rotational velocity.
}


In Case I, the thermal energy density is kept nearly constant with a very slight increase from the beginning, while the turbulent energy density is growing to the same level as the thermal energy density.
Although the magnetic energy density is amplified, it is two orders of magnitude smaller than the thermal and turbulent energies, indicating that the magnetic field has a small influence on the global energetics.

On the other hand, in Case II ({right panel of Figure \ref{fig:tevol_prsgas}}), it is found that the three different components of the energy densities finally settle down to equipartition each other. The thermal energy density, of which the initial value is about $4\ord{-11}$ erg cm$^{-3}$, rapidly decreases by an order of magnitude 
within the initial few Myr. This is due to the rapid {drop} in temperature caused by the radiative cooling in the dense {mid-plane} region. 
As a result, the gas pressure around the {mid-plane} also drops, which causes the disruption of the equilibrium configuration and excites vertical downflows to the {mid-plane}.
Thus, the turbulent (kinetic) energy density is enhanced over the thermal energy density at the early stage temporarily.
After that, the magnetic energy density is amplified {to eventually} compensate for the loss of the thermal energy {up to $t\lesssim$ 10 Myr}.
The amplification of this magnetic field is contributed by gravitational compression as mentioned above.
{Winding by differential rotation and MRI also contribute to the amplification of the magnetic fields  simultaneously.}
The figure shows that when the gradient of the total pressure consisting of the magnetic and thermal pressures can support gravity, the turbulent energy including the accretion flow settles down to be roughly comparable to the thermal and magnetic energies.

We should note that, although the equipartition among volume-averaged magnetic, thermal, and turbulent energies are satisfied in a global sense, this is not true in a local sense.
Depending on the height from the {mid-plane}, regions dominated by gas pressure or magnetic pressure are formed, which will be explained later in Section \ref{sec:multi-layer structure}.

\subsection{{$\phi$-averaged poloidal structure}}\label{sec:poloidal_structure}
\begin{figure*}
    \centering
    \plotone{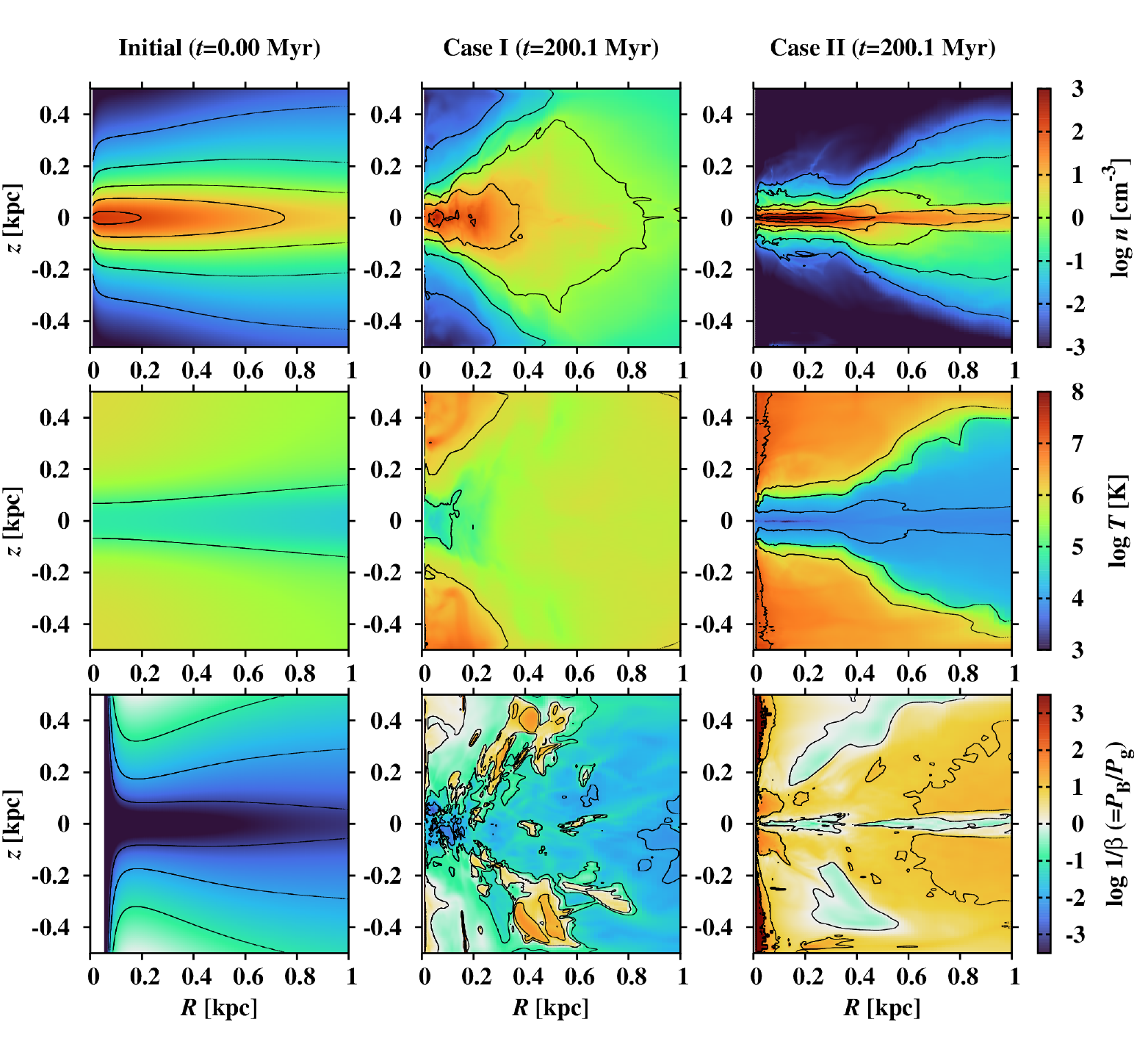}
    \caption{
    Number density $n$ (top), temperature $T$ (middle) and an inverse of plasma $\beta$ (bottom) averaged over the $\phi$ direction in a $R-z$ plane.
    The left panels present the initial profile.
    The {middle} and right panels respectively present the profile at $t=200.1$ Myr {without (Case I) and with (Case II)} cooling and heating.
    }
    \label{fig:rz_diagram_phi_averaged}
\end{figure*}

\begin{figure*}
    \centering
    \plotone{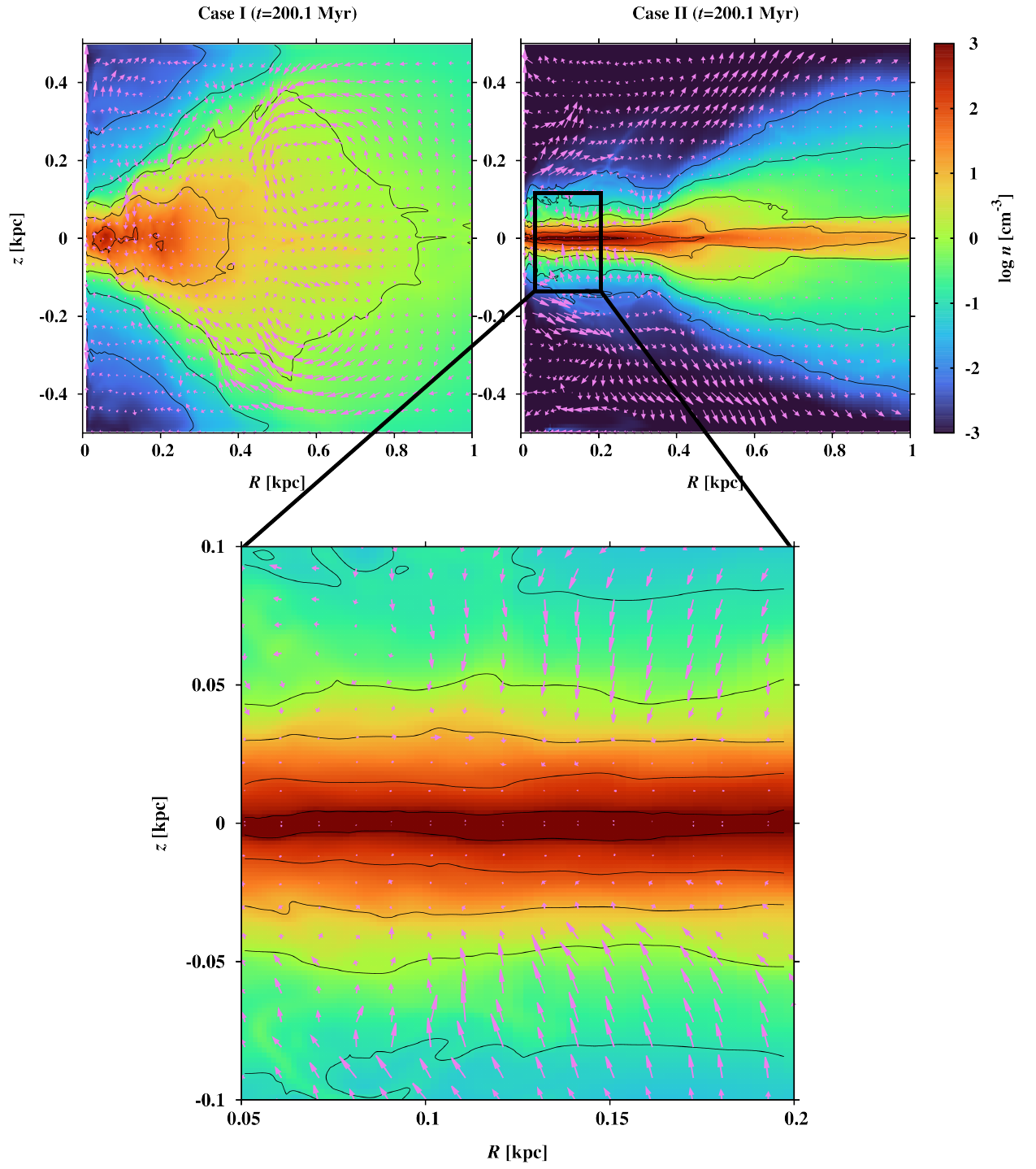}
    \caption{{Velocity fields (purple arrows) with density distribution (colors) in Case I (top left) and II (top right).
    Top two panels cover the same area as in Figure \ref{fig:rz_diagram_phi_averaged}.
    The bottom panel zooms in $0.05 < R \unit{[kpc]} < 0.2$ and $|z|<0.1$ kpc of Case II.} 
    }
    \label{fig:rz_diagram_phi_averaged_withvelocity}
\end{figure*}
Figure \ref{fig:rz_diagram_phi_averaged} shows the $\phi$-averaged density (top), temperature (middle) and inverse of plasma $\beta$ (bottom) {on a $R-z$ plane}.
The left column in the figure shows the initial distribution ($t=0.0$ Myr). 
{We note that the initial temperature $\sim 10^5$ K in the mid-plane of the bulge region is higher than realistic interstellar medium in the {mid-plane} (see Appendix \ref{sec:Appendix_B}), and therefore, the initial vertical scale height extends to about 100 pc.}
The middle and right columns show the results after time evolution of Case I and Case II at $t=200.1$ Myr.
The comparison between the top-left and top-middle panels clearly show that the gas expands vertically in Case I, as discussed Section \ref{sec:density}.
The center panel shows that the temperature at the {mid-plane} increases to
$T\approx 10^6$ K in Case I because of {magnetic heating (Section \ref{sec:magnetic_heating})}, causing the vertical expansion.

On the other hand, in the Case II,
{the initial temperature is too high in comparison with the realistic value. Hence, once the simulation started, the temperature rapidly drops around the {mid-plane} where the density is high. Therefore,} the vertical thickness at {$t = 200.1$ Myr (top-right panel)} is thinner than {that in} the initial distribution, indicating that the gas is more concentrated {near} the {mid-plane}.
In particular, the density at the {mid-plane} {in} $R<0.2$ kpc exceeds $n=10^3$ cm$^{-3}$.
It can also be seen that the inner dense region is almost flat, but the outer region ($R>0.4$ kpc) bulges vertically with increasing $R$.
As discussed below, {the vertical component of the magnetic pressure-gradient force} plays an important role in this structure.
The temperature {of the high-density gas} near the {mid-plane} is found to {be} about $T=10^{3-4}\unit{K}$.
In contrast, {low-density and} high-temperature gas of $T=10^{6-7}\unit{K}$ is distributed in the high-latitude halo regions of $|z|/R\gtrsim 1$.
This indicates the formation of multi-layer structure with a large temperature difference more than a few orders of magnitude from the {mid-plane} to the halo region like the solar atmosphere that consists of the low-temperature photosphere and chromosphere and the high-temperature corona.

{Figure \ref{fig:rz_diagram_phi_averaged_withvelocity} shows velocity field (arrows) with density (colors) on a $R-z$ diagram. It is shown from the figure that gas flows upward in the halo region in both Cases I and II. The averaged mass outflow rate in $0.05 \unit{kpc} < R <0.8 \unit{kpc}$ is $\sim$ several times $10^{-2} \unit{\Msol\ yr^{-1}}$ in Case I and several times $10^{-4} \unit{\Msol\ yr^{-1}}$ in Case II.
Additionally, one can clearly see failed winds \citep[e.g.,][]{Takasao2018a,Takasao2022a}, in which gas is floating up in the upper region from the {mid-plane} but falling back downward to the {mid-plane}.}

\begin{figure*}
    \centering
    \plottwo{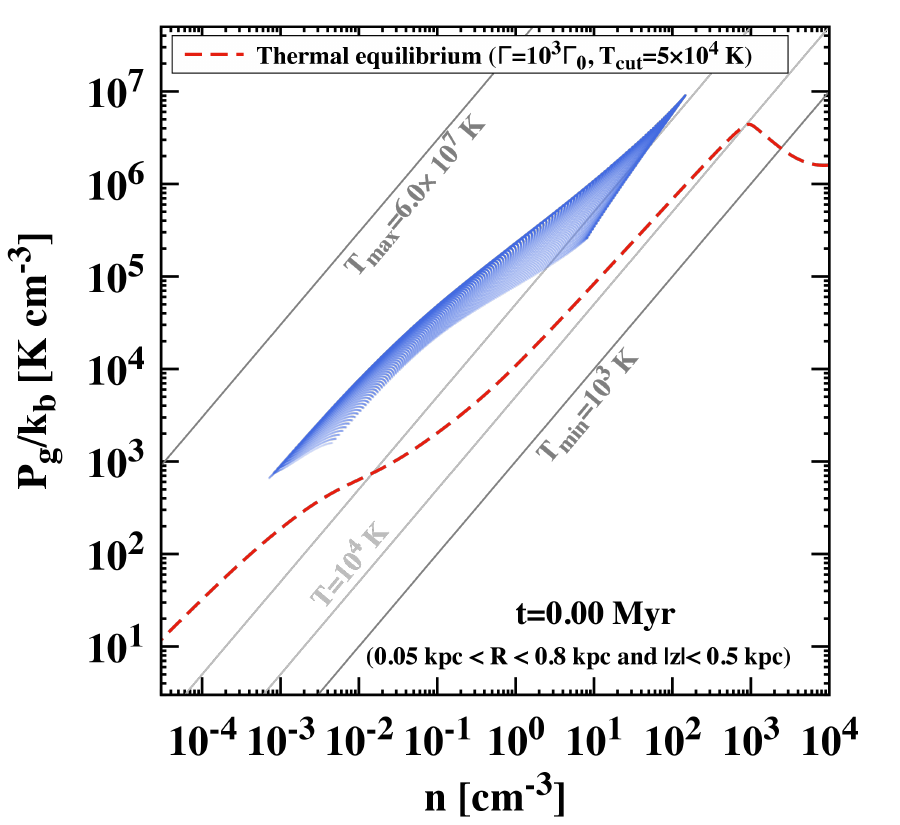}{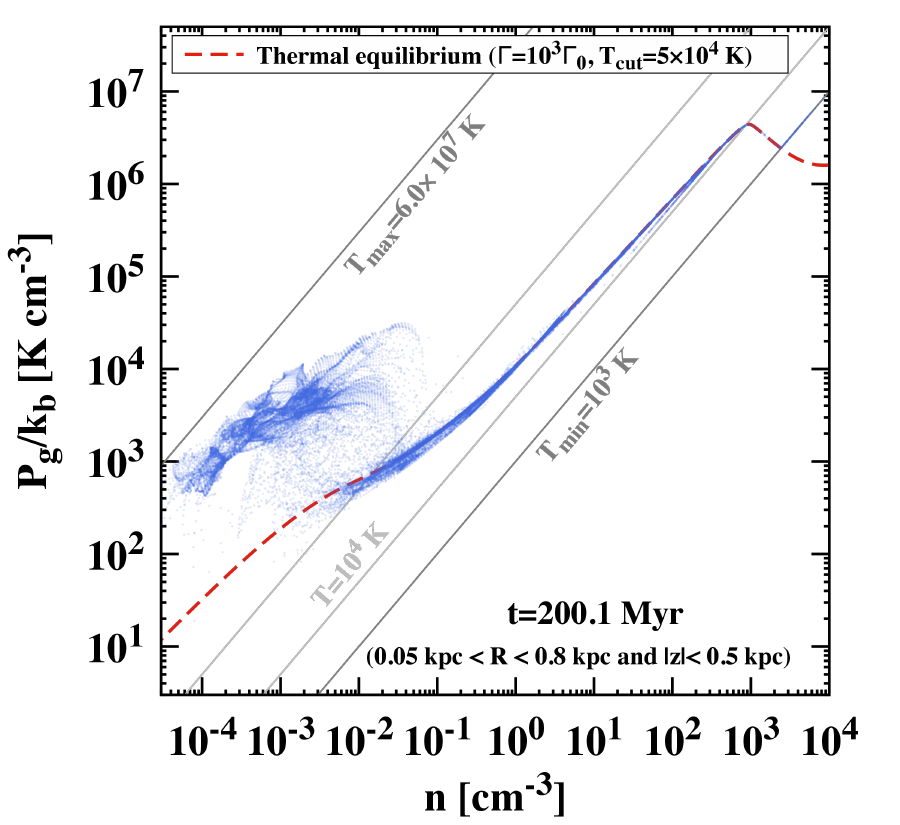}
    \caption{{$n-P$ diagram {at $t=0.0$} (left) and $t=200.1$ Myr (right). The blue points
    {are taken from the grid points} in 0.05 kpc $< R <$ 0.8 kpc and $|z|<0.45$ kpc.
    The gray solid straight lines represent the isothermal lines with $T=10^3\ (T_{\rm min}),\ 10^4$ and $6\ord{7}$ K ($T_{\rm max}$). The red dotted line is thermal equilibrium curve in the GB region ($\Gamma=10^3\Gamma_0$).} }
    \label{fig:npgraph}
\end{figure*}

As we have already mentioned, the effect of cooling and heating plays a significant role in determining these density and temperature distributions.
Figure \ref{fig:npgraph} shows $n-P$ {diagrams in a range of} $0.05 \unit{kpc} < R <0.8 \unit{kpc}$ and $|z|<0.45 \unit{kpc}$ for Case II (with radiative cooling and heating).
The left and {right panels respectively show the initial condition and} the results at $t=200.1 \unit{Myr}$.
Note that each point {in both panels} is illustrated by a translucent circle; a more opaque region with higher concentration corresponds to a larger volume fraction.
The three solid straight lines represent 
{the isothermal gas with $T=10^3$, $10^4$, and $6\ord{7}$ K} on the $n-P$ diagram, respectively.
The red dotted line plots the thermal equilibrium curve for the heating rate inside the bulge within $R<1$ kpc ($\Gamma=10^3\Gamma_0$ and the cut-off temperature of heating $T_{\rm cut}=5\ord{4}$ K; see Section \ref{sec:cooling_and_heating})

The left panel of Figure \ref{fig:npgraph} indicates that the initial temperature is higher than the thermal equilibrium temperature (red line) because the initial distribution is not in a thermally equilibrium state although it satisfies the {dynamical} balance.
Since the radiative cooling is more effective for denser gas (equation (\ref{eqn:cooling_and_heating})), the gas near {the {mid-plane}} cools down more rapidly to the thermal equilibrium temperature in a relatively short time.
In fact, the results after time evolution (right panel in Figure \ref{fig:npgraph}) show that 
the high-density gas near the {mid-plane} is distributed along the thermal equilibrium curve.
The pressure also decreases with temperature (equation (\ref{eqn_state})),
within a short period of time, especially in {the} high-density regions close to the {mid-plane} where the net cooling rate is large.
As a result, the initial equilibrium {state} is broken so that the gravity cannot be supported by the gas pressure. Downflows are excited, which further increases the density near the {mid-plane}, resulting in a more cooling-dominated environment.

In contrast, the net cooling rate is smaller in the halo region since the gas density is relatively {low}.
Therefore, the change in gas temperature in this region is expected to be moderate compared with that near the {mid-plane}.
{On the contrary}, the temperature in the halo region {increases from} the initial {value}, as seen in both Figure \ref{fig:rz_diagram_phi_averaged} and Figure \ref{fig:npgraph}.
A possible reason for the increase in the temperature {is that} there is the contribution of magnetic heating, which will be discussed in more detail in section \ref{sec:magnetic_heating}. 
In summary, the efficiency of radiative cooling is substantially affected by the gas density in the gravitationally stratified structure, resulting in the large variety in the temperature as shown in these figures.

Next, we study the evolution of magnetic field strength by examining the inverse of plasma $\beta$ in the bottom panels of Figure \ref{fig:rz_diagram_phi_averaged}.
The initial field strength is weak $< 1\unit{\mu G}$ (Equation \ref{eqn_initial_magnetic_field}), and  hence, $1/\beta$ is far below unity in the simulation domain with $< 10^{-3}$ on the {mid-plane}; the magnetic pressure is almost negligible initially in the simulation region.
Although the magnetic field strengths in Cases I (middle-bottom panel) and II (right-bottom panel) are both monotonically amplified with time, these two cases {exhibit} considerably different distributions of $1/\beta$.
In Case I, the gas pressure still remains dominant compared to the magnetic pressure in most of the GB region, although $1/\beta$ increases with the amplification of the magnetic field (Section \ref{sec:tevol_B}).

Conversely, in Case II, the magnetic pressure dominates the gas pressure in a large portion of the simulation domain.
As shown in Section \ref{sec:tevol_B}, the average saturated field strengths in Cases I and II are similar each other. Therefore, the difference in $1/\beta$ between these two cases is primarily due to the difference in the gas pressure.
As explained previously, the gas pressure rapidly decreases by the radiative cooling at the early time in Case II.
The drop in the temperature is more pronounced near the {mid-plane} where the initial density is higher.
We actually found that the average thermal energy densities in Cases I and II differ over an order of magnitude in {Section} \ref{sec:energy_density}.
Although the temperature in the halo region is kept high in both cases, the density decreases rapidly because of the infall of the gas toward the {mid-plane}, leading to the drop of the gas pressure there.
Therefore, the magnetic field plays a significant role in the dynamics and energetics of the halo gas in Case II.

However, at the {mid-plane} ($z=0.0$ kpc), the density increases locally owing to the compression by the downflow{s}. Since the temperature has reached the thermal equilibrium temperature of the warm neutral medium phase {(right panel of Figure \ref{fig:npgraph}),} the gas pressure increases as the density increases. In fact, the plasma $\beta$ at the {mid-plane} is slightly larger than unity{
; the vertical distribution of the gas is primarily supported by the gas pressure with the subdominant contribution from the magnetic pressure.}
On the other hand, the magnetic pressure is more than 10 times higher than the gas pressure in the region slightly above the {mid-plane}, hereafter we call {a} ``mid-latitude low-$\beta$ region'', which we explain in detail below.

\subsection{{Physical properties of multi-layer} structure} \label{sec:multi-layer structure}

\begin{figure*}
    \centering
    \plotone{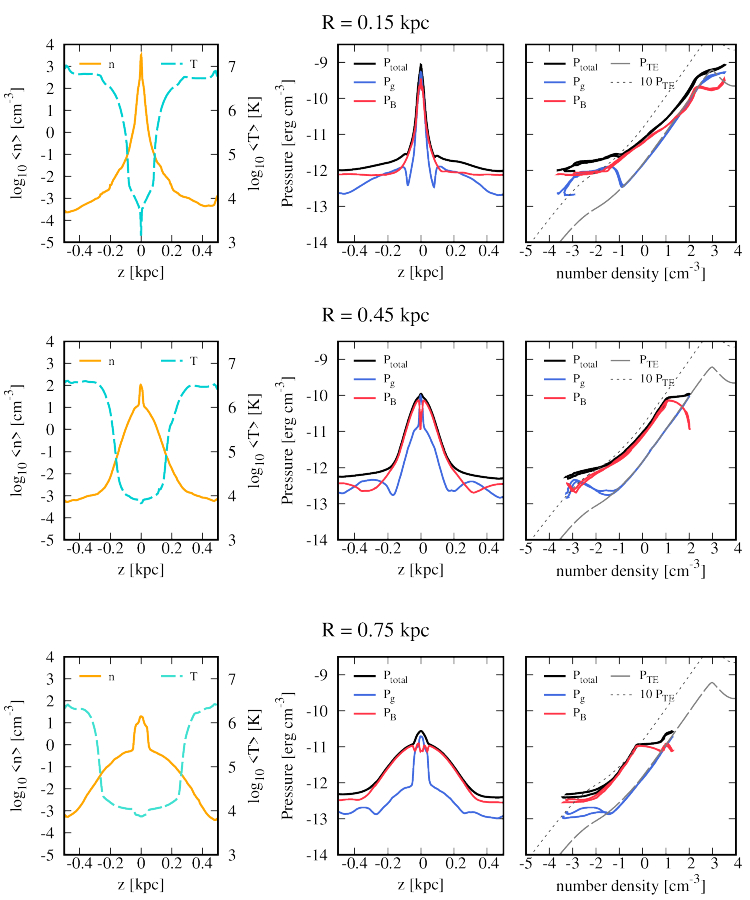}
    \caption{{
    Vertical distributions of various physical quantities at $R = 0.15$ kpc (top), 0.45 kpc (middle), and 0.75 kpc (bottom) of Case II.
    The left panels present number density (solid orange) and temperature (light blue dashed). The middle panels compare total pressure (black solid), gas pressure (blue solid), and magnetic pressure (red solid). The right panels show $n-P$ diagrams; the same line types are used for pressure; the thermal equilibrium curved for $P_{\rm TE}$ (gray dashed) and 10 $P_{\rm TE}$ (gray dotted) are also plotted (see text for the detail).}
    }
    \label{fig:z_profile_Rdl}
\end{figure*}

It is found that {in} Case II, while 
the gas pressure dominates the magnetic pressure on the {{mid-plane}}, 
the magnetic pressure is dominant in the upper layers.
Figure \ref{fig:z_profile_Rdl} compares the vertical profiles of various physical quantities averaged over time between 184.8 Myr and 203.3 Myr at $R=0.15$ (top), 0.45 (middle), and 0.75 (bottom) kpc for Case II.
The left panels show the {vertical} profile of {number} density, $n$ (yellow {solid} line), and temperature, $T$ ({sky-blue} dotted line).
The middle {panels compare the vertical distributions of} gas pressure, $P_{g}$ ({blue} line), magnetic pressure, $P_{B}$ ({red} line), and total pressure, $P_t (=P_{g}+P_{B}$; black line). In the right panels, the different components of the pressure are presented against $n$.
In these $n-P$ diagrams, the location close to the {mid-plane} corresponds to the larger $n$ (right) side.

First, we discuss the result at $R=0.45$ kpc (middle row of Figure \ref{fig:z_profile_Rdl}). The high-density region with $n\gtrsim 10$ cm$^{-3}$ is formed across the {mid-plane} with the vertical thickness of $|z|\lesssim\pm$ {a} few 10 pc by the initial infall of gas from the upper regions (Section \ref{sec:poloidal_structure}). The {force} balance is sustained between the downward gravity and the gas pressure, which dominates the magnetic pressure. However, the gas pressure rapidly decreases with increasing $|z|$ 
so that in the upper layer of $|z|\gtrsim$ a few 10 pc, the gravity is mainly supported by the magnetic pressure. From an energetics point of view, the thermal energy reduced by the radiative cooling (Section \ref{sec:energy_density}) is partially compensated by the amplified magnetic energy, which forms the mid-latitude low-$\beta$ region. Interestingly, similar low-$\beta$ regions are obtained in a global MHD simulation for disks of spiral galaxies \citep{Wibking2023a} and active galactic nuclei \citep{Kudoh20a}.
The further upper halo region is occupied by low-density ($n\lesssim 10^{-2}$ cm$^{-3}$) and high-temperature ($T>10^6$ K) gas.

These features can also be well captured in the $n-P$ diagram (middle-right panel in the Figure \ref{fig:z_profile_Rdl}). The gas pressure almost exactly follows the thermal equilibrium relation in the high density part, $n\gtrsim 0.1$ cm$^{-3}$. Although the magnetic pressure is subdominant in the Galactic-plane region ($n\gtrsim 10$ cm$^{-3}$), it greatly dominates the gas pressure in most of the upper layers. An interesting point is that the plots of $P_{B}$ and $P_{g}$ are almost in parallel with keeping $P_B/P_g(=1/\beta)\approx 10$ in the mid-latitude low-$\beta$ region with $10^{-1.5}$ cm$^{-3}$ $\lesssim n\lesssim 10$ cm$^{-3}$. Since $P_g$ almost coincides with the thermally equilibrium pressure, $P_{\rm TE}$, in this region, $P_B$ follows the line of $10P_{\rm TE}$. In the further lower density part, $n\lesssim 10^{-1.5}$cm$^{-3}$, which corresponds to the halo region, the gas pressure, $P_g$ diviates upward from $P_{\rm TE}$. This indicates that the halo gas is heated by the dissipation of magnetic energy, which is discussed in Section \ref{sec:magnetic_heating} (see also Figure \ref{fig:rz_diagram_phi_averaged}).

While the qualitative properties at the inner ($R=0.15$ kpc; top row of Figure \ref{fig:z_profile_Rdl}) and outer ($R=0.75$ kpc; bottom row) locations are basically the same as those at $R=0.45$ kpc, the vertical thickness moderately increases with $R$ because of the weakend gravity. An additional remark for the inner location is that the downward gravity is not fully supported by the gas and magnetic pressures there. Therefore, gas continues to fall down from the halo region ({Figure \ref{fig:rz_diagram_phi_averaged_withvelocity}}).

\section{Discussions}
\subsection{Thickness of low-$\beta$ region} \label{sec:thickness-beta} 
 \begin{figure*}
     \centering
     \plotone{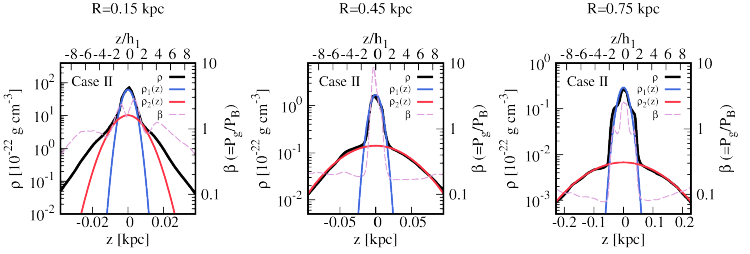}
     \caption{Density distribution in Figure \ref{fig:z_profile_Rdl} {zoomed near the mid-plane. The black solid line is the density of the numerical simulation of Case II. The red and green solid lines are the density profiles, supported by gas pressure, $\rho_1$ (equation \ref{eqn_rho1_profile}), and magnetic pressure, $\rho_2$ (equation \ref{eqn_rho2_profile}), respectively.}
     %
     The {purple dashed} line represents the plasma $\beta$ profile.
     {The top axis is normalized by gas scale height, $h_1$.}}
     \label{fig:magnetic_scale_height}
 \end{figure*}

We have shown that the magnetically dominated regions are formed in the upper layers above the {mid-plane} (Section \ref{sec:multi-layer structure}).
We investigate the detailed properties and formation mechanism of these mid-latitude low-$\beta$ regions by inspecting the vertical distributions of density and magnetic field.
Figure \ref{fig:magnetic_scale_height} shows the vertical profile of density (black  solid line) and plasma $\beta$ ({pink dotted} line) {in the $z$ direction} at the same Galactic center distances $R=0.15, 0.45, 0.75$ kpc as in Figure \ref{fig:z_profile_Rdl}.
%

{When the vertical gravity is expanded around $z=0$, the leading-order term gives}
\begin{align}
g_z\simeq \left. \di{g_z}{z}\right |_{z=0}\ z \equiv \omega^2 z, \label{eqn_g_z}
\end{align}
{where, we note that $\omega$ becomes the Keplerian angular velocity for point-source gravity.
When the vertical gravity is supported by the gas pressure and vertical motion can be neglected, equation (\ref{eqn_forcebalnce_z}) gives vertical density distribution. Applying equation (\ref{eqn_g_z}), we obtain 
}
\begin{align}
\frac{1}{\rho}\pa{\rho}{z}=-\frac{\omega^2}{c_s^2}z, \label{eqn_balance_z}
\end{align}
{where we assumed an isothermal equation of state, $P_{\rm g}=\rho c_s^2$, for a constant sound velocity. The solution of equation (\ref{eqn_balance_z}) is }
%
%
\begin{align}
\rho_1 (z) = \rho_0 \exp{(-z^2/h_1^2)},\label{eqn_rho1_profile}
\end{align}
{where $h_1=c_s/\omega$ is the gas pressure scale height.}
{The density, $\rho_0$, and the sound velocity, $c_s$, at the {mid-plane} are evaluated from the numerical data; $\omega$, is calculated from equation \ref{eqn_gravity_all}}.
Following the same procedure, {in the magnetically dominant condition}, we derive
\begin{align}
 \rho_2 (z) = \rho'_0 \exp{(-z^2/h_2^2)},\label{eqn_rho2_profile}
\end{align}
where $h_2=v_A/{\omega}$ and $v_A=|B|_{{h}}/\sqrt{4 \pi \rho'_0}$. 
{Here, $|B_h|_{\rm max}$ is the maximum horizontal field strength in the low-$\beta$ regions for a given $R$ {in the numerical data.}}
Here, $\rho'_0$ is the normalization density that reproduces the density distribution in the upper regions.

\begin{table*}
    \centering
    \begin{tabularx}{0.75\linewidth}{c|c|c|c|c|c|c|c}
        \hline
        $R$ & $\rho_0$ &
        $c_s$ & $\omega$ &
        $h_1$ &$\rho_0{'}$  & $|B|_{z,\rm max}$  & $h_2$ \\
        (kpc) & ($10^{-22} \unit{g\ cm}^{-3}$) & (km s$^{-1}$) & 
        ($10^{-14}$ s$^{-1}$) & (pc) & ($10^{-22}\unit{g\ cm}^{-3}$) &
        ($\mu$G) & (pc) \\
        \hline
        0.15 & 62.20 & 3.47  & {2.65} & {4.2} & 10.15 & 74.36 & 10.0 \\
        0.45 & 1.70 & 7.72 & {1.83} & {13.6} & 0.17 & 45.88 & 57.0\\
        0.75 & 0.30 & 8.72 & {1.14} & {24.7} & 6.77 & 17.17 & 162.0 \\
        \hline
    \end{tabularx}
    \caption{The value used to estimate scale height at each radius.}
    \label{table:scale_height}
\end{table*}

The {blue and red} solid lines in Figure \ref{fig:magnetic_scale_height} indicate $\rho_1(z)$ and $\rho_2(z)$ for each $R$.
{The specific values used to find the scale height at each radius are shown in the Table \ref{table:scale_height}.}
From the figure, the density distribution in the $z$ direction can be 
{well fitted by the sum of the two components,} 
$\rho_1(z)$ and $\rho_2(z)$.
The density distribution near the {mid-plane} ($|z/h_1|<2$) is almost consistent with $\rho_1(z)$.
This is because the density distribution near the {mid-plane} is mainly supported by the gas pressure {so that} its thickness is determined by $h_1$. 
{In contrast, the magnetic pressure dominates in the upper regions.}
Thus, the gas is spread out vertically beyond $z=h_1$.
At $R=0.45$ kpc (the middle of Figure \ref{fig:magnetic_scale_height}), the thickness of the gas density in the upper region agrees well with the density distribution of $h_2$ determined by the magnetic pressure. 
{The same tendency is obtained for the density distribution at $R = 0.75$ kpc.}
On the other hand, at $R = 0.15$ kpc, the density distribution {extends beyond}
the thickness $h_2$ determined by magnetic pressure. 
{Therefore, the gas located above $|z|\gtrsim 10$ pc cannot be supported by magnetic or gas pressure but still continuously falls downward}

 \begin{figure*}
     \centering
    \plotone{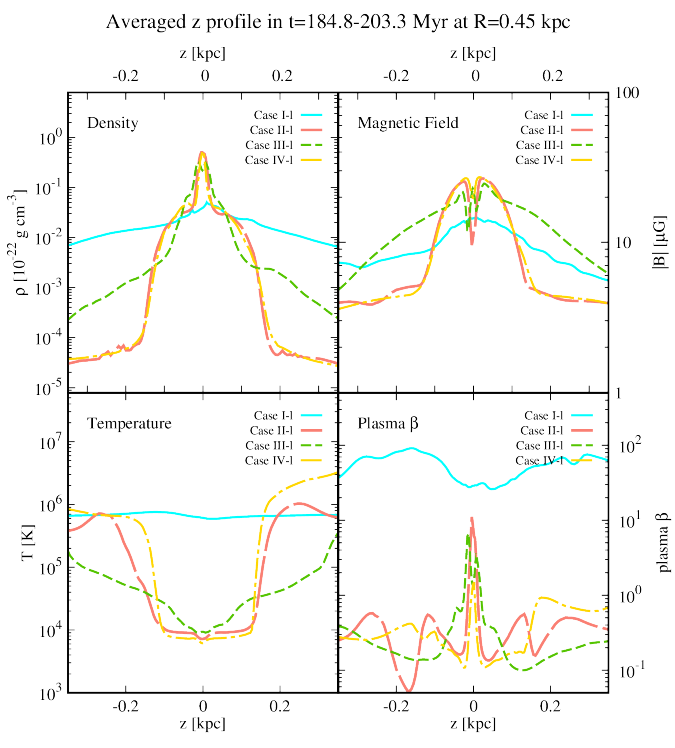}
     \caption{
     Comparison of {vertical profiles for different heating rates}. Each panel represents {vertical distributions at} $R = 0.45 $kpc {of} density (top left), magnetic field strength (top right), temperature (bottom left), and plasma $\beta$ values (bottom right). 
     The lines correspond to the results of {Case I-l} ({adiabatic case; {cyan} solid}), {Case II-l} ({fiducial cooling and heating; {pale red} dashed}), {Case III-l} ({reduced net cooling; green dotted}), and {Case IV-l} ({enhanced net cooling; yellow dot-dashed}), respectively.
    }
     \label{fig:magnetic_scale_height_compared}
 \end{figure*}

The presence of the mid-latitude low-$\beta$ regions is a characteristic outcome of the radiative cooling.
Our treatment for the radiative cooling and heating is an approximated one (Section \ref{sec:cooling_and_heating}). Thus, we investigate the impact of the magnitude of the net cooling rate on the density distribution with 3D MHD simulations by changing the radiative heating rate. In {Case III-l (IV-l)} we increase (decrease) the radiative heating rate, $\Gamma$, to 10 (1/10) of the original setting in Case II.  Since the increase (decrease) of the heating corresponds to the reduction (enhancement) of the net cooling, we call {Case III-l (IV-l)} the case with reduced (enhanced) cooling.

{As described in Section \ref{sec:setup}, in Cases III-l and IV-l}, we employ half the numerical resolution compared to the fiducial models of Cases I and II. 
{
The low resolution simulations are performed for Cases I and II
\footnote{{While Cases I-l and II-l give slightly lower density around the mid-plane than the original Cases I and II, the overall trends are not so different (see Appendix \ref{sec:Appendix_C}).
}}.
We compare these four cases (Cases I-l - IV-l) with the same resolution.
}


{Figure \ref{fig:tevol_B} indicates the volume averaged magnetic field strengths in $0.05 \unit{kpc} < R < 0.8 \unit{kpc}$ and $|z|<0.45 \unit{kpc}$ are very similar in Cases II-l, III-l and IV-l.} 
Figure \ref{fig:magnetic_scale_height_compared} {compares the vertical distributions of Cases {I-l, II-l, III-l and IV-l} at $R = 0.45$ kpc.}
{
The qualitative vertical distributions of {Cases III-l and IV-l} remain unchanged even under the different net cooling rates; low-temperature and high-density gas around the midplane is sandwiched by mid-latitude low-$\beta$ regions 
(see the lower right panel of Figure \ref{fig:magnetic_scale_height_compared}).
However, the difference in the net cooling significantly influences in the upper region away from the {mid-plane}.
In the reduced net cooling case ({Case III-l}), it is found that the density (top left), magnetic field (top right), and temperature (left bottom) structures exhibit intermediate characteristics between {Case I-l and Case II-l}.
In the enhanced net cooling case ({Case IV-l}), the structure is similar to {Case II-l}. 
}

In the {mid-plane} where the gas density is high, the cooling rate is significantly dominant over the heating rate, even though the heating rate is increased by a factor of 10 as done in {Case III-l}.
Therefore, the net cooling rate is almost unaffected there and the system reaches a thermal equilibrium state within a short time. On the other hand, in the upper region with lower density, the cooling rate decreases, and thus, the difference in heating rate has a significant impact on the net cooling rate. As a result, the difference in the heating rate affects the cooling time compared to the dynamical time, leading to the observed difference in the vertical distribution in the upper regions.

These results indicate that the height at which magnetic pressure dominates ($\beta<1.0$) {is controlled by the adopted {net cooling} rate}.
{This is} because the thermal equilibrium temperature, which
{depends on the net cooling rate, determines the gas pressure.}
In our simulations, we {are using the simplified treatment}
for the radiative cooling and heating.
In particular, we have assumed the heating rate $\Gamma_0$ for a typical interstellar cloud near the solar system in equation (\ref{eqn_ht-Gamma}).
{In reality, photoelectric heating and} heating by X-rays and cosmic-rays depend on the surrounding environment in different ways.
{Also, we are not solving the ionization states of atoms that are crucially important in determining radiative cooling.
More realistic simulations with these effects should be carried out in our future works.}

\subsection{Distribution of magnetic field strength}\label{sec:discuss_CNM}
  \begin{figure}
      \centering
      \plotone{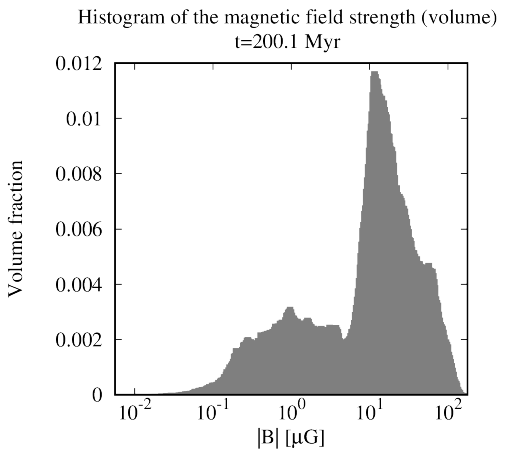}
      \plotone{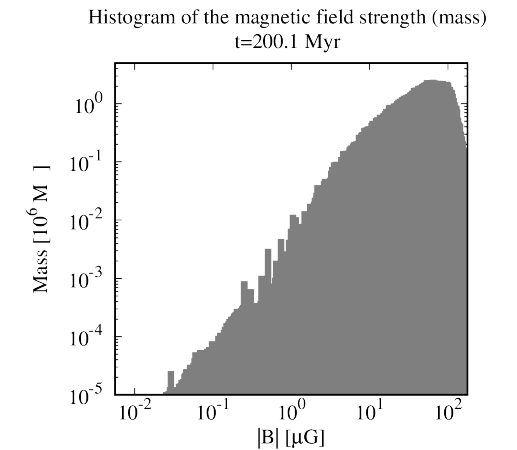}
      \caption{
      {Histograms of the volume fraction (top) and mass (bottom) against magnetic field strength. Each bin is spacing via $\Delta \ln{B} = 0.02$.}}
      \label{fig:hist_bfd_inCMZ}
  \end{figure}

 \begin{figure}
     \centering
     \plotone{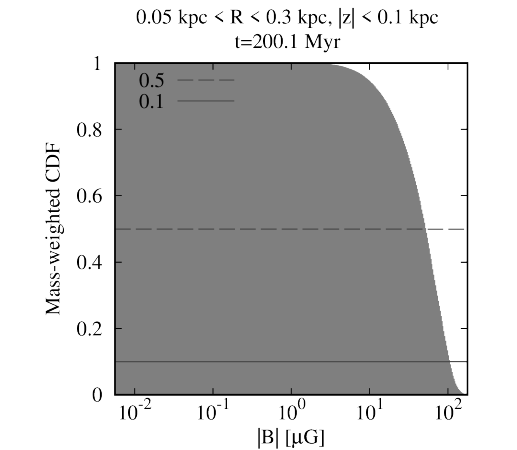}
     \caption{{Cumulative mass for magnetic field strength.
     The integration is taken from the maximum $|B|_{max} = 200$ $\mu$G to smaller $|B|$.
     The horizontal solid and dashed lines indicate fraction of 0.1 and 0.5, respectively.}}
     \label{fig:CDF_bfd_inCMZ}
 \end{figure}

 {In our MHD simulations, the magnetic field strength is amplified to $> 10 \unit{\mu G}$ with the maximum strength of a few to several $100 \unit{\mu G}$, which is consistent with the observationally suggested value, $10-1000 \unit{\mu G}$ in the GB region \citep{Morris1996,Pillai15a,Crocker10a}. In this subsection, we further examine the distribution of the obtained magnetic field from a statistical viewpoint, particular focusing on the dependence on density.}

Figure \ref{fig:hist_bfd_inCMZ} shows the histograms 
as a function of the magnetic field strength {in $0.05\unit{kpc}<R< 0.3\unit{kpc}$, which corresponds to the scale of the CMZ.}
The volume fraction {(the top panel of Figure \ref{fig:hist_bfd_inCMZ})} is dominated by low-density gas above the {mid-plane}.
Since the gravity {is} stronger {at small} $R$, 
the gas that was cooled by radiative cooling cannot maintain the {dynamical} equilibrium.
Therefore, low-density gas falls down toward the {mid-plane} and forms a compressed dense region.
The two peaks in the volume fraction diagram 
{correspond to the compressed high-density gas near and on}
the {mid-plane} and low-density gas that falls from the upper layers.
From the volume fraction {histogram}, the spatially averaged magnetic field strength is estimated to be about 10 $\mu$G. 


{The bottom panel of Figure 17 presents the mass histogram. In each $i$-th bin spacing with $\Delta \ln B=0.02$, the mass in the cells that satisfy $\ln{B_i} < \ln{B} < \ln{B_i}+\Delta \ln{B}$ is integrated:}
\begin{align}
\Sigma^{\ln Bi+\Delta \ln B}_{\ln B_i}\rho \Delta V
\rightarrow\int^{\ln Bi+\Delta \ln B}_{\ln B_i}\rho dV.
\end{align}
{The mass spectrum can focus on denser regions; therefore, the mass weighted average} magnetic field strength is higher than the {volume weighted} field strength, with the peak close to 100 $\mu$G.
In particular, the mass fraction that exceeds 100 $\mu$G is about 11\% (see in Figure \ref{fig:CDF_bfd_inCMZ}).

In the inner part of the GB region, there are dense clouds with strong magnetic fields where the field strength locally exceeds 1 mG. For one specific example, \citet{Pillai15a} estimated to be magnetic field strengths of 1-2 mG inside the "Brick" region, a high-density dark nebula in the CMZ.
Such strong-field regions tend to be correlated with high-density molecular clouds.
The total mass of CMZ at $R<300$ pc is $M_{\rm CMZ}=2-6\ord{7}M_{\rm \odot}$ \citep{Dahmen98a,Ferriere2007}.
The average molecular gas density in the CMZ is $\rho=10^{-20}$ g cm$^{-3}$ \citep{Guesten83a,Bally87a,mills18a}.
Therefore, the volume of molecular gas in the Galactic center region is estimated to be $V=M_{\rm CMZ}/\rho\sim7\ord{60}$ cm$^3$, which is about 1\% of the total volume {in} $R<300$pc {and} $|z|<100$ pc.
On the other hand, the molecular gas is likely to be the major component {in mass} since it accounts for more than 90\% of the total mass of the entire CMZ scale.
The current numerical setting cannot handle such high-density and low-temperature clouds because of the the cut-off temperature for the radiative cooling (equation (\ref{eqn:cooling_and_heating})) and the insufficient numerical resolution. In our future study, we plan to investigate strong-field concentrations in high-density regions by improving the numerical treatment. 

\subsection{Cosmic-Rays and $\gamma$-Ray Excess in the GB region}
In this paper, we show that the magnetic field supports the density structure in the $z$ direction in the upper region of the {mid-plane}. On the other hand, it is suggested that cosmic-ray pressure may contribute to the same extent \citep{Boulares90a}.
It has also been reported that cosmic-rays can be a source of heating of dilute ionized gas and drive outflow \citep{Shimoda22a}. The study of dilute gas at higher altitudes ($z>1-10 \unit{kpc}$) should incorporate the effects of cosmic-rays and supernova explosions, which are not considered in this study.

The cosmic-rays around the {GB} region are also related to
the observed $\sim 100$~TeV $\gamma$-ray excess \citep{HESS_Collabo2016}.
The cosmic-ray protons with an energy of $\sim$~PeV can be responsible for
the $\sim 100$~TeV $\gamma$-ray emissions although their origin and acceleration
process are still under debated \citep[e.g.,][]{Shimoda2022a}. In the most accepted scenario of cosmic-ray
acceleration, the diffusive shock acceleration at supernova remnant shocks, the maximum energy of accelerated
particles is determined by the magnetic filed strength at the shock upstream \citep[e.g.,][]{Lagage1983a,Lagage1983b,Bell2004a}.
The required field strength is $>100~\mu$G for accelerating $\sim$PeV cosmic-rays at the supernova remnant shock.
{If the volume fraction of gas with such a strong field was large}, the PeV cosmic-ray acceleration would occur
everywhere in the GB region. Thus, further investigation of the field strength
distribution is a necessary step to elucidate the origin of the high energy $\gamma$-rays
and cosmic-rays.

\subsection{Magnetic heating} \label{sec:magnetic_heating}

In Figures \ref{fig:npgraph} and \ref{fig:z_profile_Rdl}, {we explained}
that the low-density gas in {the halo region} is significantly heated from the thermal equilibrium temperature to a high-temperature state. 
{We here discuss the role of the dissipation of magnetic fields in heating the halo gas.}

The magnetic energy can be transformed to kinetic and thermal energy via magnetic reconnection processes that happen in the geometrically thin current sheets where we cannot neglect physical resistivity. In realistic situations, it is regarded that magnetic turbulence induces small-scale reconnections \citep{Lazarian99a,Lazarion20a}, which finally converts the magnetic energy to the thermal energy of the surrounding gas.
In general, such {reconnections of turbulent magnetic fields} take place at very fine scales, which requires extremely small sizes of numerical cells.
{It is not realistic to perform such high-resolution simulations in our global treatment. Instead, utilizing the energy conserved scheme we are employing (Section \ref{sec:basic_equations}), we estimate the heating rate that is expected from the dissipation of magnetic energy \citep{Suzuki2006a, Matsumoto14a}.}

In our simulations dissipative phenomena are treated by the MHD Riemann solver; the physical heating by viscosity within the sub-grid scale is
handled by {this shock capturing} scheme.
Although {this} method cannot describe {and directly capture} the details of the physical dissipation scale, we believe that it can 
{give a reasonable estimate on the global magnetic heating rate.}

The magnetic energy conservation law is expressed as
\begin{align}
    \pa{}{t}\enc{\frac{B^2}{8\pi}}
    -\frac{1}{4\pi}\vectorb{\nabla}\cdot\encl{(\vectorb{v}\cdot\vectorb{B})\vectorb{B}-B^2\vectorb{v}}\nonumber\\
    -\vectorb{v}\cdot\encl{\vectorb{\nabla}\enc{\frac{B^2}{8\pi}} -\frac{1}{4\pi}(\vectorb{B}\cdot\vectorb{\nabla})\vectorb{B}}
    =-\frac{J^2}{\sigma_c},
    \label{eqn_mgntE}
\end{align}
where the term on the {right-hand} side {denotes} Joule heating {rate} using the {electric current} $J$ and the electric conductivity $\sigma_c$.
{We note that,} since $\sigma_c=\infty$ is assumed in the ideal MHD,
{the} right-hand side is zero in the ideal MHD limit.
However, {what we are solving in our simulations is not equation (\ref{eqn_mgntE}) but equation (\ref{eqn_induction}) for the conservation of the total energy. Therefore, the sum of the three terms on the left-hand-side of equation (\ref{eqn_mgntE}) is not zero numerically, which gives a finite value of the {Joule} heating rate.}
{This non-zero Joule heating at each numerical cell is counted in the change of the thermal energy.}
This increase in {the} thermal energy due to {the} numerical dissipation can be interpreted as {the energy released at} MHD shocks and {by} small-scale reconnections as a consequence of turbulent cascade that occurs at a sub-grid scale.
In the following {discussion}, 
{the Joule heating rate, $J^2/\sigma$, derived {numerically} from equation (\ref{eqn_mgntE})} is referred to as the ``magnetic heating rate''.

\begin{figure}[t]
    \centering
    \plotone{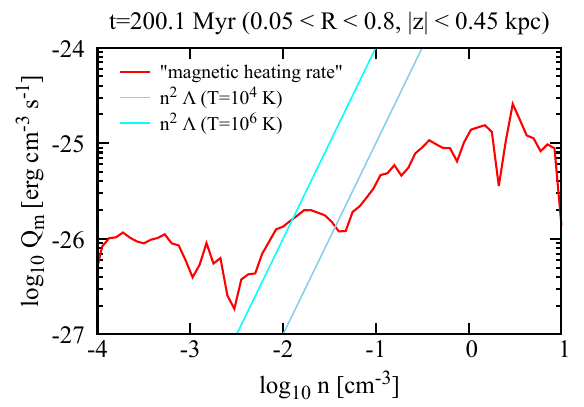}
    \caption{{
    Volume average magnetic heating rate for number density.
    }}
    \label{fig:mgnht}
\end{figure}

Figure \ref{fig:mgnht} shows magnetic heating rate (red solid line) averaged in the range of $0.05\unit{kpc}<R<0.8\unit{kpc}$ and $|z|<0.45$ kpc at $t=$200.1 Myr, in comparison to the radiative cooling rates for different temperatures (solid bluish lines).
Although the magnetic heating is largely dominated by the radiative cooling in the high-density regions with $n\gtrsim 1$cm$^{-3}$, it is not negligible in lower-density regions. In particular, for $n<10^{-2}$ cm$^{-3}$, the magnetic heating rate {exceeds} the radiative cooling rate,
causing the deviation of the temperature in the halo region from the equilibrium value shown in Figures \ref{fig:npgraph} and \ref{fig:z_profile_Rdl}.

Although we are taking account for radiative cooling and heating in the energy equation, we do not consider the effect of thermal conduction (diffusion). The timescale{, $\tau_i$,} of the thermal conduction for fully ionized gas in the {GB} region can be estimated {as follows}: 

\footnotesize
\begin{align}
\tau_i=\frac{\rho L^2}{\kappa_{i}}
\simeq 3000 \enc{\frac{n}{10^{-3}\unit{cm}^{-3}}}
\enc{\frac{L}{100 \unit{pc}}}^2\enc{\frac{kT}{0.5 \unit{keV}}}^{-5/2} \unit{yr}, \label{eqn_thermal_conduction_i}
\end{align}
\normalsize
where $\kappa_i=1.24\ord{-6}\ T^{5/2}\unit{[erg\ K^{-1}\ cm^{-1} s^{-1}]}$ is the thermal conductivity \citep{Parker1953a,Asai07a}.
This estimate indicates that thermal conduction is more effective in lower-density and higher-temperature environments, where magnetic heating is also actively working.
If we substitute typical temperature, $T=10^6 - 10^7$K, and density, $n < 10^{-2}$ cm$^{-3}$, into equation (\ref{eqn_thermal_conduction_i}), we obtain $\tau_i$ ranging from $10^{-4}$ Myr to 0.1 Myr.
Since this is much shorter than the rotation time $\sim$ a few 10 Myr in the GB region, it is expected that the temperature of the hot gas should have a flatter temperature profile by thermal conduction than that shown in Figure \ref{fig:rz_diagram_phi_averaged}.

On the other hand, magnetic heating and radiative heating are weaker against radiative cooling in the high-density regions near the {mid-plane}. Thus, the averaged gas temperature is low $\lesssim$ $10^4 \unit{K}$ {there}. 
In such low-temperature gas, neutral atoms contribute to the thermal conduction. The corresponding conductive timescale is

\footnotesize
\begin{align}
    \tau_n=\frac{\rho L^2}{\kappa_{n}}
\simeq 10^{10} \enc{\frac{n}{10^{3}\unit{cm}^{-3}}}
\enc{\frac{L}{100 \unit{pc}}}^2\enc{\frac{kT}{0.1 \unit{keV}}}^{-5/2} \unit{yr}, \label{eqn_thermal_conduction_n}
\end{align}
\normalsize
where $\kappa_n=2.5\ord{3}\ T^{1/2}\unit{[erg\ K^{-1}\ cm^{-1} s^{-1}]}$ \citep{Parker1953a,Koyama00a}.
%
%
Therefore, the effect of thermal conduction is negligible on the scale of $L \gtrsim 10 - 100\unit{pc}$ near the {mid-plane}.

\subsection{{Application to Milky Way -- the stellar bar}}
{
Our simulations are focusing on the GB region of general spiral galaxies with axisymmetric bulge structure (e.g., M31, NGC 278, NGC 628\footnote{{We note that recent star formation activity is weak in NGC 628 \citep{Hoyer2023a}, and then, this galaxy in the current epoch may not be a good target for our simulations.}}, NGC 772 and NGC 4030).}
{If we are to compare to the Milky way, we have to take into account the effect of the bar potential. This is because }
the stellar bar structure with a size of about 3 kpc in our galaxy is widely known from infrared observations \citep{Matsumoto82a,Nakada1991,Weinberg1992,Dwek1995}.
How the gravitational field affects the distribution of the interstellar gas and its motion inside this bar structure has been discussed for more than 30 years.
Assuming the inner bar potential based on the basic theory of \citet{Binney91a}, several intersecting orbits are formed, and the orbits near the Galactic center, called X1 and X2 orbits, reproduce the {elongated} orbits of the gas in and around the CMZ and the gas accretion flow along the dust lanes \citep{Athanassoula92a}.
The orbits and structures of the interstellar gas in the bar {potential} have been examined in detail by N-body simulation \citep{Rodriguez-Fernandez08a}.
In recent years, high-resolution hydrodynamic simulations provide quantitative results, and the impact on the bar potential on the star formation history is being discussed \citep{Sormani18a, Armillotta19a,Baba2020a,Tress2020a}. Indeed, it is identified that the contribution of {the stellar bar} forms a stream along dust lanes {with} the time-averaged inflow rate {of} a few $M_\odot$ yr$^{-1}$ \citep{Sormani19a}.

It is also discussed how the bar potential contributes to the gas distribution and motion via the magnetic field in the Galactic center region \citep{kim2012a}.
%
Recently, \citet{Moon23a} performed 3D MHD numerical simulations {under the magnetized inflowing gas along a dust lane into the Galactic center region, which is expected from the stellar bar.}
%
{They reported that the amplified magnetic field plays a substantial role in the suppression of star formation in the nuclear ring region and the excitation of inflows from the ring to the nuclear disk.}
In our next step, we are also planning to include the effect of the stellar {bar} in global MHD simulations.

\section{Conclusion}

We examined the impact of magnetic fields on gas dynamics in the GB region using 3D-MHD simulations with radiative cooling and heating.
Although the thermal properties are greatly affected by the radiative cooling, the saturated field strengths are not so different in the cases with and without the radiative effects; the magnetic fields are amplified from the initial value $< \unit{\mu G}$ to the volume integrated average of $\approx 10 \unit{\mu G}$ with several ${\rm \mu G}$ in localized field concentrations. 
We also found rough equipartition between the volume-averaged magnetic and thermal energies, whereas the local plasma $\beta$ value varies with height.
This result emphasizes the substantial role of magnetic fields in controlling dynamical and thermal properties in the GB region.

An important finding is that, by the effect of the radiative cooling in combination of magnetic fields, {a} vertically multi-layered structure is created. The cool dense gas that is mainly supported by gas pressure is distributed in a thin layer on and near the {mid-plane}. Above that, low-$\beta$ plasma occupies the vertical range of $z=50-150 \unit{pc}$. The magnetic pressure, which largely dominates the gas pressure in the mid-latitude low-$\beta$ regions, mainly maintains this vertical thickness comparable to the observed scale height of neutral atomic gas. The further upper regions are occupied by low-density and high-temperature halo gas.

In the cool dense {mid-plane} and the mid-latitude low-$\beta$ regions, the temperature is determined by the thermal equilibrium between the radiative cooling and heating. In contrast, the temperature is largely deviated from the equilibrium value in the Galactic halo. This indicates that the magnetic heating {may be} important in such low-density gas. 



\begin{acknowledgments}
K. K. thanks Y. Fukui, R. Enokiya and A. Tanikawa for fruitful discussions.
This work was supported by JSPS KAKENHI Grant Nos. 19J11052 (K. K.),
17H01105, 21H00033, 22H01263 (T. K. Suzuki) and 18H05436, 18H05437 (S. Inutsuka),
and by Program for Promoting Research on the Supercomputer Fugaku by the RIKEN Centre for Computational Science (Toward a unified view of the universe: from large-scale structures to planets, grant 20351188 – PI J. Makino) from the MEXT of Japan.
This research used computational resources of Oakforest-PACS provided by Multidisciplinary Cooperative Research Program in Center for Computational Sciences, University of Tsukuba, Yukawa-21 at YITP in Kyoto University and Cray XC50 at Center for Computational Astrophysics, National Astronomical Observatory of Japan.
\end{acknowledgments}



\appendix

\section{Fitting function for radiative cooling} \label{sec:fitting_lamda}
In the Figure \ref{fig:ini_T2mu}(a), the black points show the data table of \citet{Sutherland93a}.
We used the fitting function of the fourth order equation created from the table data from \citet{Sutherland93a} as $\Lambda_\textrm{h}$ at the high temperature side of $T>10^4\unit{K}$:
\begin{align}
    \Lambda_\textrm{h}(T)=c_0 + c_1 (\log_{10}{T})
    +c_2 (\log_{10}{T})^2+c_3 (\log_{10}{T})^3
    +c_4 (\log_{10}{T})^4,
\end{align}
where $c_0=-156.919$, $c_1=84.2271$, $c_2=-19.0317$, $c_3=1.85211$, $c_4=-0.0658615$.

\section{Initial distribution} \label{sec:Appendix_B}
{We set up the initial distributions, following \citet{Suzuki15a}.}
%
%
{First, we determine the temperature distribution (the middle left of Figure \ref{fig:rz_diagram_phi_averaged}). We give $T=1.25\ord{5}K$ at the {mid-plane} in the bulge, $T=5 \ord{3} \unit{K}$ at the {mid-plane} in the disk, $T=2.0\ord{6} \unit{K}$ at $z=0.5$ kpc in the bulge, and $T=2.0\ord{5} \unit{K}$ $z=0.5$ kpc in the disk. We smoothly interpolate temperature in the boundary layers between these four regions.
}



{
Next, the density at the {mid-plane} ($z=0.0$ kpc) is assumed to be 0.05 times the stellar mass distribution that is determined by $\Phi_{\rm *}$ (equation \ref{eqn_gravity} and Section \ref{sec:potential}).
We integrate equation \ref{eqn_forcebalnce_z} to determine $P_g (R,z)$, using $g_z=-\pa{\Phi}{z}$. As temperature is already fixed, the density distribution $\rho(R,z)$ is derived from $P_g$.
Finally, by integrating equation (\ref{eqn_forcebalnce_R}) along the radial direction, we determine the initial rotation frequency.
}


\section{Comparison of results for different resolutions}\label{sec:Appendix_C}
\begin{figure*}
     \centering
    \plotone{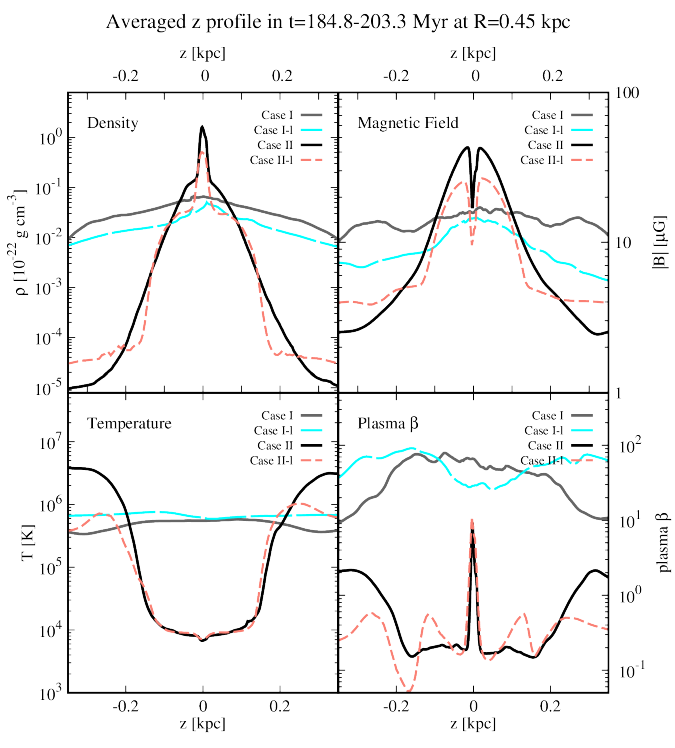}
     \caption{
     {Same as Figure 16 but for Cases I (gray solid), II (black solid), I-l ({cyan} dashed), and II-l ({pale red} dashed).}
    }
     \label{fig:magnetic_scale_height_compared_resolution}
 \end{figure*}

{
Figure \ref{fig:magnetic_scale_height_compared_resolution} 
compares the vertical distributions of various physical quantities taken from Cases I, II, I-l, and II-l.
The top left panel shows that density around the {mid-plane} is slightly higher in the high-resolution case than in the low-resolution case.
The magnetic field strength (top right) is also slightly more amplified near the mid-plane plane in the high-resolution case, compared with the low-resolution case.
On the other hand, temperature (bottom left) in the mid-plane region does not depend on numerical resolution.
This is because the thermal equilibrium temperature determined by radiative cooling is {not affected by numerical} resolution. 
However, temperature in the halo region is higher for higher resolution owing to the resolution dependence of magnetic heating.
}


\bibliography{ref,library}
\bibliographystyle{aasjournal}



\end{document}